\documentclass[review]{elsarticle}
\usepackage{amsmath}
\usepackage{amsthm}
\usepackage{amssymb}
\usepackage{titlesec}
\usepackage[thinlines,thiklines]{easybmat}
\newcommand{\cent}{{|}\!\!\mathrm{c}}
\usepackage{graphicx}
\theoremstyle{definition}\newtheorem{Df}{Definition}
\theoremstyle{plain}\newtheorem{Th}{Theorem}
\theoremstyle{definition}\newtheorem{Rm}{Remark}
\theoremstyle{definition}
\theoremstyle{plain}\newtheorem{Pp}[Th]{Proposition}
\theoremstyle{plain}\newtheorem{Co}[Th]{Corollary}
\theoremstyle{plain}\newtheorem{Lm}[Th]{Lemma}

\begin{document}
\begin{frontmatter}
\title{ \Large{Characterizations of   one-way general quantum finite automata\tnoteref{t}}}
\tnotetext[t]{This work is supported in
part by the National Natural Science Foundation (Nos.
60873055, 61073054), the
Natural Science Foundation of Guangdong Province of China (No. 10251027501000004), the Fundamental Research Funds for the Central Universities (No. 10lgzd12), the Program
for New Century Excellent Talents in University (NCET) of China,
and the project of  SQIG at IT, funded by FCT and EU FEDER projects
Quantlog
POCI/MAT/55796/2004 and
QSec PTDC/EIA/67661/2006, IT Project QuantTel, NoE Euro-NF, and the SQIG LAP initiative. Lvzhou Li is partly supported  by the China Postdoctoral Science Foundation funded project (20090460808).}
\author[a1]{Lvzhou Li}
\author[a1,a2,a3]{Daowen Qiu\corref{one}}
\author[a1]{Xiangfu Zou}
\author[a1]{ Lvjun Li}
\author[a1]{ Lihua Wu}
\author[a2]{Paulo Mateus}

 \cortext[one]{Corresponding author.\\
 \indent {\it E-mail
address:} issqdw@mail.sysu.edu.cn (D. Qiu), lilvzhou@gmail.com (L.
Li).}

\address[a1]{Department of
Computer Science, Sun Yat-sen University, Guangzhou 510006,
  China}

\address[a2]{SQIG--Instituto de Telecomunica\c{c}\~{o}es, Departamento de Matem\'{a}tica, Instituto Superior T\'{e}cnico,
Universidade T\'{e}cnica de Lisboa,
 Av. Rovisco Pais 1049-001, Lisbon, Portugal}

 \address[a3]{The State Key Laboratory of Computer Science, Institute of Software, Chinese  Academy of Sciences,
 Beijing 100080, China}

\begin{abstract}

Generally, unitary transformations limit the computational power of
{\it quantum finite automata }(QFA). In  this paper we study a
generalized model named {\it one-way general quantum finite
automata} (1gQFA), in which each symbol in the input alphabet
induces a trace-preserving quantum operation, instead of a unitary
transformation. Two different kinds of 1gQFA will be studied: {\it
measure-once one-way general quantum finite automata} (MO-1gQFA)
where a measurement deciding to accept or reject is performed at
the end of a computation, and {\it measure-many one-way general
quantum finite automata} (MM-1gQFA) where  a similar measurement
is performed after each trace-preserving quantum operation on
reading each input symbol.

We characterize the measure-once model from three aspects: the
closure property, the language recognition power, and the
equivalence problem. Specially, we prove that MO-1gQFA recognize, with bounded error, precisely the set of all
regular languages. Our results imply that some quantum finite automata
proposed in the literature, which were expected to be more
powerful, still can not recognize non-regular languages.

 We prove that MM-1gQFA also recognize only regular languages with
 bounded error. Thus,  MM-1gQFA and MO-1gQFA have the same language recognition
power, which is greatly different from the conventional case in which the number of times the measurement is performed in the
computation generally affects the language recognition power of one-way QFA.
Finally, we present  a  sufficient and necessary condition for  two MM-1gQFA to be  equivalent.

\end{abstract}
\end{frontmatter}

\section{ Introduction}

Since Shor's quantum algorithm for factoring integers in polynomial time
\cite{Shor} and Grover's algorithm of searching in database of
size $n$ with only $O(\sqrt{n})$ accesses \cite{Grover}, quantum computation and information has  attracted  more and more attention in the community. As we know, these algorithms are based on {\it quantum Turing machines}
which seem complicated to realize using today's experiment
technology. Therefore, since making powerful quantum computers is still a long
term goal, it is important to study ``small-size" quantum processors (such as {\it quantum finite automata}) using variations of the models that have shown their relevance in the classical cases \cite{Gru99}.

Indeed,
quantum finite automata (QFA), as a theoretical model for quantum
computers with finite memory,  have interested many researchers
(see, e.g., [1-15, 18-19, 21-28, 30-36, 41-42]). From the
theoretical point of view, exploring QFA may redound to getting an
insight into the power of quantum computation.

So far, several models of QFA have been studied. These models
differ from each other mainly by two factors: the moving direction
of the tape head and the measurement times. Roughly speaking, we
have two kinds of QFA: {\it one-way} QFA (1QFA) where the tape
heads are allowed only to move towards right at each step and {\it
two-way} QFA (2QFA) where the tape heads are allowed to move
towards left or right, and even to stay stationary.

 The model of 2QFA was firstly studied by Kondacs and Watrous
\cite{KW97}. It was proved that 2QFA can  not only
recognize\setcounter{footnote}{0}\footnote{Without additional
explanation, in this paper, recognizing a language always means
recognizing a language with bounded error.} all regular languages,
but also recognize some non-regular languages such as
$L_{eq}=\{a^{n}b^{n}|n>0\}$ in linear time \cite{KW97}. It is worth pointing
out that any two-way probabilistic automaton needs exponential time
to recognize a non-regular language \cite{KF91}. Therefore, 2QFA are more powerful than
the classical counterparts. However, it
seems to be difficult to implement a 2QFA, since the size of a
2QFA's quantum part depends on the length of the input.

Compared with 2QFA, 1QFA seem simpler to design and implement.
In general, theoretical problems regarding 1QFA are easier to
investigate than those regarding 2QFA. Thus, 1QFA may be an
appropriate beginning of the study on quantum computing models.
Two important types of 1QFA are {\it measure-once } 1QFA (MO-1QFA)
proposed by Moore and Crutchfield \cite{MC00} where a measurement
is performed at the end of a computation, and {\it measure-many }
1QFA (MM-1QFA) defined by Kondacs and Watrous \cite{KW97} where a
measurement is performed at each step during a computation. It has
been proved that both MO-1QFA and MM-1QFA can  recognize only a
proper subset of regular languages. More exactly, MO-1QFA
recognize only group languages \cite{BC01,BP99}, and MM-1QFA
recognize more languages than MO-1QFA but can not recognize all
regular languages such as $L=\{a,b\}^*a$ \cite{KW97}.

Obviously,  both MO-1QFA and MM-1QFA have a very limited
computational power. Thus, some generalizations or modifications
were made to the definition of 1QFA, with the expectation to
enhance the computational power of 1QFA. An usual modification to
the definition of 1QFA is to allow an arbitrary projective
measurement as an intermediate step in the computation. Note that
in the conventional 1QFA such as MO-1QFA and MM-1QFA, the
measurement operation is allowed only at deciding ``accept'' or
``reject''. At the same time, we know that the operation of
measurement plays an important role in quantum information
processing. Thus, it is natural  to allow the projective
measurement as an intermediate step in the computation of QFA.

For instance, Ambainis {\it et al } \cite{Amb06} studied the
so-called {\it Latvian} QFA (LQFA) in which the operation corresponding
to an input symbol is the combination of a unitary transformation
and a projective measurement.  LQFA can be regraded as a
generalized version of MO-1QFA. In fact, LQFA are closely related with the classical model PRA-C \cite{MK02}, and from the results in \cite{Amb06} it follows that the two models recognize the same class of languages, i.e., the languages whose syntactic monoid is a block group  \cite{Amb06}.  More concretely,  a language is recognized by LQFA if and only if the language is a Boolean combination of languages of the form $L_0a_1L_1\dots a_kL_k$ where the $a_i$'s are letters and the $L_i$'s are group languages.
 Hence, LQFA can  recognize only a proper subset of  regular languages; for example, LQFA can not recognize regular languages $\Sigma^*a$ and $a\Sigma^*$ \cite{Amb06}.
 In fact, the class of languages recognized by LQFA is a proper subset of the languages recognized by MM-1QFA, since that on one hand   MM-1QFA can recognize all the languages  recognized by LQFA \cite{Amb06}, and on the other hand,   MM-1QFA can recognize language $a\Sigma^*$ that can not be recognized by LQFA\setcounter{footnote}{1}\footnote{For example, to recognize $a\{a,b\}^*$, we construct an MM-1QFA with initial state $|q_0\rangle$, reject state $|q_{rej}\rangle$ and accepting state $|q_{acc}\rangle$, and the transition operators are constructed as:
 \begin{align*}
&U_a(|q_0\rangle)=
\frac{1}{\sqrt{2}}|q_1\rangle+\frac{1}{\sqrt{2}}|q_{acc}\rangle,
~~U_a(|q_1\rangle)=
\frac{1}{\sqrt{2}}|q_1\rangle-\frac{1}{\sqrt{2}}|q_{acc}\rangle,\\
&U_b(|q_1\rangle)=\frac{1}{\sqrt{2}}|q_1\rangle+\frac{1}{\sqrt{2}}|q_{acc}\rangle,
~~U_b(|q_0\rangle)=|q_{rej}\rangle,
~~U_\$(|q_0\rangle)=|q_{acc}\rangle,~\hspace{2mm}U_\$(|q_1\rangle)=|q_{rej}\rangle,
\end{align*}
where some transitions that are not important have not been described.}.

The measure-many version of LQFA  was defined by Nayak \cite{Nay99}, and we call this model as GQFA in this paper. It can be seen that  GQFA allow more general operations than theses models mentioned before. From the results about LQFA stated above, it follows that GQFA are strictly more powerful than LQFA. However, GQFA still can not recognize all regular languages; for example, GQFA can not recognize language $\{a,b\}^*a$ \cite{Nay99}. Also, it is still not known whether GQFA can recognize strictly more languages than MM-1QFA. An interesting result about GQFA is that there exist languages for which GQFA
take exponentially more states than those of the corresponding
classical automata  \cite{ANT02,Nay99}.

Bertoni {\it et al} \cite{Ber03} defined a
model called {\it one-way quantum finite automata with control
language} (CL-1QFA). The accepting behavior of this model is greatly different from the ones of these models stated before, which is controlled by  the sequence of the  measurement result obtained
at each step in the computation. If the result sequence is in a given language, then the input is accepted. In Ref. \cite{MP06}, it was
proved that CL-1QFA recognize exactly regular languages with
bounded error. Recently, Qiu and Mateus {\it et al } \cite{QZM09} studied {\it one-way quantum finite automata
together with classical states} and showed that this model can also recognize any regular language with no error.

Besides these 1QFA  mentioned above, there are some other models of 1QFA
which  go more further  in the direction
of modifying the original definition  of 1QFA. For instance, Ciamarra
\cite{Cia01} thought that the reason for the computational power
of 1QFA being weaker than that of their classical counterparts is
that the definition of 1QFA neglects the concept of quantum
reversibility. Thus, following the idea in Bennett \cite{Ben73},
Ciamarra \cite{Cia01} proposed a new model of 1QFA that was
believed to be strictly reversible, and whose computational power
was proved to be  at least equal to that of (one-way) classical
automata. Paschen \cite{Pas00}  introduced another model of 1QFA
named {\it ancilla QFA}, where an ancilla quantum part is
imported, and then the internal control states and the states of
the ancilla part together evolve by a unitary transformation.
Paschen \cite{Pas00} showed that  ancilla QFA can recognize any
regular language with certainty.

The two models of QFA stated above can recognize at least regular
languages. In a certain sense, the increased computational power
is from the generalization of the operations allowed by
the  models. In fact, the two QFA defined in \cite{Cia01} and in
\cite{Pas00} have a common point, that is, the internal state controller
 together with some auxiliary quantum parts evolve by a
unitary transformation, and thus the evolution of the internal
control states is generally not unitary. Therefore, a natural
question is: how much computational power can no-unitary
operations bring to quantum finite automata?

We will address the above question in this paper. For that, we
study the generalized version of 1QFA, called {\it one-way general
quantum finite automata} (1gQFA), in which each symbol in the
input alphabet induces a trace-preserving quantum operation
instead of a unitary transformation. Two kinds of 1gQFA will be
studied:   {\it measure-once one-way general quantum finite
automata} (MO-1gQFA) which can be seen as a generalized version of  MO-1QFA, and  {\it measure-many one-way
general quantum finite automata} (MM-1gQFA), a generalized version
of  MM-1QFA.

We study MO-1gQFA  from three aspects: the closure property,
the language recognition power, and the equivalence problem.  In
fact, such a kind of QFA has already been proposed by Hirvensalo
\cite{Hir08}, but with no further attention paid to this model.
Hirvensalo \cite{Hir08} showed that MO-1gQFA can simulate any
probabilistic automaton, and thus can recognize any regular
language. In general, it is believed that the unitarity of
evolution  puts some limit on the computational power of QFA. Now,
MO-1gQFA allow any physical admissible operation---the
trace-preserving quantum operation. Then MO-1gQFA are expected to
be more powerful. Specially, we have such a question: can MO-1gQFA
recognize some non-regular languages?

In this paper,  we will prove that despite the most general
operations allowed, MO-1gQFA  can recognize only regular
languages with bounded error. Moreover, the two types of QFA
defined in \cite{Cia01} and in \cite{Pas00} are shown to be within
the model of MO-1gQFA, and thus recognize only regular languages.
Another problem worthwhile to be pursued is the equivalence between MO-1gQFA.
We will give a sufficient and necessary condition for two MO-1gQFA to be equivalent.
Also, we will  present some closure properties of MO-1gQFA.

We study MM-1gQFA from two aspects: the language recognition power
and the equivalence problem. Generally, the number of times the measurement is performed in the
computation affects the computational power of
1QFA. For instance, MM-1QFA can recognize more languages than
MO-1QFA, and GQFA also recognize more languages than LQFA. Therefore, it is expected that MM-1gQFA are more powerful than MO-1gQFA. However, we will prove that MM-1gQFA  also recognize
only regular languages with bounded error. Thus, MM-1gQFA and
MO-1gQFA have the same computational power. This reveals an
essential difference between 1QFA and their generalized
versions. Finally, we discuss the equivalence problem of MM-1gQFA.
Specifically, we give a sufficient and necessary condition to
determine whether two MM-1gQFA are equivalent or not. This also
offers a different solution to the equivalence problem of MM-1QFA
discussed in Li and Qiu \cite{LQ08a}.

 It is worth pointing out that all the above discussions regarding
 MM-1gQFA are  based on such  a result proved by us: an MM-1gQFA can be
 simulated by a relaxed version of MO-1gQFA whose operation
 corresponding
 to the input symbol is a general linear super-operator, not necessarily a
 trace-preserving quantum operation.

\section{Preliminaries}

\subsection{Notations and quantum operations}
 Some  notations used in this paper are explained here. $|S|$ denotes the cardinality of set $S$.
 For non-empty set $\Sigma$, by $\Sigma^{*}$ we mean the
set of all strings over $\Sigma$ with finite length. $|w|$ denotes the
length of string $w$. Symbols $*$, $\dagger$, and $\top$  denote
the conjugate operation, the conjugate-transpose operation, and
the transpose operation, respectively. $\text{Tr}(A)$ denotes the
trace of matrix (operator) $A$.  $supp(A)$ denotes the support of
operator $A$. For a positive operator $A$, $supp(A)$ is the space
spanned by the eigenvectors of $A$ corresponding to the no-zero
eigenvalues. $\dim V$ denotes the dimension of finite-dimensional
space $V$. Generally, we use ${\cal H}$ to denote a
finite-dimensional Hilbert space.  Let $L({\cal H})$ denote the
set of all linear operators from ${\cal H}$ to itself. A mapping
$\Phi$ in this form: $L({\cal H})\rightarrow L({\cal H})$ is
called a super-operator on ${\cal H}$.

Let $\mathbb{C}$ and $\mathbb{R}$ denote the sets of complex
numbers and real numbers, respectively.  Let $\mathbb{C}^{n\times
m}$ denote the set of all $n\times m$ complex matrices. For two
matrices $A\in\mathbb{C}^{n\times m}$ and $B\in\mathbb{C}^{p\times
q}$, their direct sum is defined as
 $$A\oplus B=\left[%
\begin{array}{cc}
  A & O \\
  O & B \\
\end{array}%
\right],$$ and their tensor product is
$$A\otimes B=\begin{bmatrix}
  A_{11}B & \dots & A_{1m}B \\
  \vdots & \ddots & \vdots \\
  A_{n1}B &\dots & A_{nm}B \
\end{bmatrix}.$$

The detailed  background on quantum information can be referred to
 \cite{NC00}, and here we just introduce some
notions. According to the postulates of quantum mechanics, the
state of a closed quantum system is represented by a unit vector
$|\psi\rangle$ in a Hilbert space ${\cal H}$, and the state
evolution of a closed quantum system is described by a unitary
transformation on ${\cal H}$.  A more general tool to describe the
state of a quantum system is the density operator.  A density
operator $\rho$ on Hilbert space ${\cal H}$ is a linear operator
satisfying the following conditions:
\begin{itemize}
    \item [(1)] (Trace condition)  $\rho$ has trace equal to $1$, that is,  $\text{Tr}(\rho)=1$.
     \item [(2)] (Positivity condition) $\rho\geq 0$, that is, for any
     $|\psi\rangle \in {\cal H}$, $\langle\psi|\rho|\psi\rangle\geq
     0$.
\end{itemize}
By $D({\cal H})$ we mean the set of all density operators on
Hilbert space ${\cal H}$.

In practice, an absolutely closed system does not exist, because a
system interacts more or less with its outer environment, and thus
it is open. Then the state evolution of an open quantum system is
characterized by a {\it quantum operation} \cite{NC00}. A quantum
operation, denoted by ${\cal E}$, has an {\it operator-sum
representation} as
\begin{align}
{\cal E}(\rho)=\sum_kE_k\rho E_k^\dagger,\label{OP}
\end{align}
where $\rho$ is a density operator on the input space ${\cal H}_{in}$, ${\cal
E}(\rho)$ is a linear operator on the output space ${\cal H}_{out}$, and the set of
$\{E_k\}$ known as {\it operation elements} are linear operators
from ${\cal H}_{in}$ to ${\cal H}_{out}$. Furthermore, ${\cal E}$
is said to be trace-preserving if the following holds:
\begin{align}
\sum_kE_k^\dagger E_k=I,
\end{align}
where $I$ is the identity operator on ${\cal H}_{in}$.

 Any physical admissible operation is a trace-preserving quantum
operation (also called a completely positive trace-preserving
mapping), which  has another representation---{\it Stinespring
 representation}:
 \begin{align}
 {\cal E}(\rho)={\text Tr}_{{\cal H}_a}(V\rho V^\dagger),\label{SR-operation}
 \end{align}
where $V$ is a linear isometry operator from ${\cal H}_{in}$ to
${\cal H}_{out}\otimes {\cal H}_a$, and ${\text Tr}_{{\cal H}_a}$
is the operation of partial trace  that discards the subsystem
$a$.

When the input space and the output space of quantum operation ${\cal E}$ are the same, say ${\cal H}$,
we say ${\cal E}$ is a quantum operation acting on ${\cal H}$. In fact, the quantum operations used in the subsequent
sections are all in this case.
\subsection{A brief review on  MO-1QFA and  MM-1QFA} In this
paper, we are interested in  quantum finite automata with a
one-way tap head. Two important models of 1QFA are MO-1QFA firstly
defined by Moore and Crutchfield \cite{MC00} and MM-1QFA proposed
by Kondacs and Watrous \cite{KW97}. For the readers having a good
understanding on the general models studied in this paper,
we first present a brief review on MO-1QFA and MM-1QFA in the
following.

An MO-1QFA is defined as a quintuple ${\cal A}=(Q, \Sigma,
|\psi_{0}\rangle,\{U(\sigma)\}_{\sigma\in\Sigma},Q_{acc})$, where
$Q$ is a set of finite states, $|\psi_{0}\rangle$ is the initial
state that is a superposition of the states in $Q$, $\Sigma$ is a
finite input alphabet, $U(\sigma)$ is a unitary transformation for
each $\sigma\in\Sigma$, and $Q_{acc}\subseteq Q$ is the set of
accepting states. The computing process of MO-1QFA ${\cal A}$ on
 input string
$x=\sigma_{1}\sigma_{2}\cdots\sigma_{n}\in\Sigma^{*}$ is as
follows:  the unitary transformations
$U(\sigma_1),U(\sigma_2),\cdots,U(\sigma_n)$ are performed in
succession on the initial state $|\psi_{0}\rangle$, and finally a
measurement is performed on the final state, deciding to accept
the input or not. The languages recognized by MO-1QFA with bounded
error are group languages \cite{BP99}, a proper subset of regular
languages.

An MM-1QFA is defined as a 6-tuple ${\cal
M}=(Q,\Sigma,|\psi_{0}\rangle,\{U(\sigma)\}_{\sigma\in\Sigma\cup
\{\cent,\$\}},Q_{acc},Q_{rej})$, where $Q,Q_{acc}\subseteq
Q,|\psi_{0}\rangle,\Sigma,\{U(\sigma)\}$ are the same as those in
the MO-1QFA defined above, $Q_{rej}\subseteq Q$ represents the set
of rejecting states, and $\cent,\$\not\in\Sigma$ are respectively
the left end-marker and the right end-marker. For any input string
$\cent x\$$ with $x\in\Sigma^{*}$, the computing process is
similar to that of MO-1QFA except that after every transition,
${\cal M}$ measures its state with respect to the three subspaces
that are spanned by the three subsets $Q_{acc}, Q_{rej}$, and
$Q_{non}$, respectively, where $Q_{non}=Q\setminus (Q_{acc}\cup
Q_{rej})$.
 The languages recognized by MM-1QFA with bounded error are more
 than those recognized by MO-1QFA, but  still a proper subset
 of regular languages.

 From the study on MO-1QFA and MM-1QFA, we make two observations: (i) the number of times the
measurement is performed in the computation affects the computational
power of 1QFA; (ii) by considering just unitary transformations one
limits the computation power of 1QFA such that the two typical models of 1QFA (MO-1QFA and MM-1QFA) are less
powerful than their classical counterparts.

Inspired by these observations  above, in this paper we are going
to study the generalized versions of MO-1QFA and MM-1QFA, in which
the most general operations---trace-preserving quantum operations
are allowed at reading each input symbol. By studying these models, we hope to address such a question: what are
the limitations imposed by unitary transformations in the computation
power of 1QFA? Or, in other words, what extra computational power can
non-unitary transformations bring to 1QFA?

\section{One-way general quantum finite automata (I): MO-1gQFA}
In this section, we consider the model of  MO-1gQFA which has a
one-way tape head, and in which each symbol in the input alphabet
induces a trace-preserving quantum operation. In the subsequent
sections, after giving the definition of MO-1gQFA, we will discuss
the closure property, the language recognition power, and the
equivalence problem for MO-1gQFA.

\subsection{Closure properties of MO-1gQFA}\label{3.1} First,we give the definition of MO-1gQFA as
follows.
\begin{Df}
An MO-1gQFA ${\cal M}$ is a five-tuple ${\cal M}=\{ {\cal
H},\Sigma,\rho_0,\{{\cal E}_\sigma\}_{\sigma\in\Sigma},
P_{acc}\}$, where ${\cal H}$ is a finite-dimensional Hilbert
space, $\Sigma$ is a finite input alphabet, $\rho_0$, the initial
state of ${\cal M}$, is a density operator on ${\cal H}$, ${\cal
E}_\sigma$ corresponding to $\sigma\in\Sigma$ is a
trace-preserving quantum operation acting on ${\cal H}$, $P_{acc}$
is a projector on the subspace called accepting subspace of ${\cal
H}$. Denote $P_{rej}=I-P_{acc}$, then $\{P_{acc}, P_{rej}\}$ form
a projective measurement on ${\cal H}$.
\end{Df}

On input word $\sigma_1\sigma_2\dots\sigma_n\in\Sigma^{*}$, the
above MO-1gQFA ${\cal M}$ proceeds as follows: the quantum
operations ${\cal E}_{\sigma_1},{\cal E}_{\sigma_2},\dots,{\cal
E}_{\sigma_n}$ are performed on $\rho_0$ in succession, and then
the projective measurement $\{P_{acc}, P_{rej}\}$ is performed on
the final state, obtaining the accepting result with a certain
probability. Thus,  MO-1gQFA  ${\cal M}$ defined above induces a
function $f_{\cal M}: \Sigma^*\rightarrow [0,1]$ as
\begin{align}
f_{\cal M}(\sigma_1\sigma_2\dots\sigma_n)=\text{Tr}(P_{acc}{\cal
E}_{\sigma_n}\circ\dots\circ{\cal E}_{\sigma_2}\circ{\cal
E}_{\sigma_1}(\rho_0)),
\end{align}
where ${\cal E}_2\circ{\cal E}_1(\rho)$ stands for  ${\cal
E}_2({\cal E}_1(\rho))$.  In fact, for every $x\in\Sigma^*$,
$f_{\cal M}(x)$ represents the probability that ${\cal M}$ accepts
$x$.

 In the following, we present some closure properties of MO-1gQFA.

\begin{Th} The class of MO-1gQFA are closed under the following
operations:

 (i) If $f$ is a function induced by an MO-1gQFA, then
$1-f$ is also induced by an MO-1gQFA.

(ii) If $f_1,f_2,\dots, f_k$ are functions induced by MO-1gQFA,
then $\sum_i^kc_if_i$ is also  induced by an MO-1gQFA for any real
constants $c_i>0$ such that $\sum_i^k c_i=1$.

(iii) If $f_1,f_2,\dots, f_k$ are functions induced by MO-1gQFA,
then $f_1f_2\cdots f_k$ defined as $f_1f_2(w)=f_1(w)f_2(w)$ is
also induced by an MO-1gQFA.\label{Closure}
\end{Th}

\begin{proof} (a) If $f$ is induced by an MO-1gQFA ${\cal
M}$ with projector $P_{acc}$, then $1-f$ can be induced by
MO-1gQFA ${\cal M}^{'}$ that is almost the same as ${\cal M}$, but
with a projector $I-P_{acc}$.

(b) We prove item (ii) for $k=2$ in detail. Assume that $f_i$ is
induced by MO-1gQFA ${\cal M}_i=\{ {\cal
H}_i,\Sigma,\rho^{(i)}_0,\{{\cal
E}^{(i)}_\sigma\}_{\sigma\in\Sigma}, P^{(i)}_{acc}\}$ with
$i=1,2$, respectively. Then for $c_1, c_2$ satisfying $c_1>0$,
$c_2>0$ and $c_1+c_2=1$, we construct ${\cal H}={\cal H}_1\oplus
{\cal H}_2$, and $\rho_0=c_1\rho^{(1)}_0\oplus c_2\rho^{(2)}_0$.
$\rho_0$ is obviously a density operator on ${\cal H}$. Moreover,
for every $\sigma\in\Sigma$, we construct ${\cal E}_\sigma={\cal
E}^{(1)}_\sigma\oplus {\cal E}^{(2)}_\sigma$;  more specifically,
if ${\cal E}^{(1)}_\sigma$ and ${\cal E}^{(2)}_\sigma$ have
operator element sets $\{E_i\}_{i\in N}$ and $\{F_j\}_{j\in M}$,
respectively,  then ${\cal E}_\sigma$ is constructed such that it
has operator element set $\{\frac{1}{\sqrt{M}}E_i\oplus
\frac{1}{\sqrt{N}}F_j\}_{i\in N,j\in M}$. Then we have
\begin{align}
&\sum_{i\in N,j\in M}\left(\frac{1}{\sqrt{M}}E_i\oplus
\frac{1}{\sqrt{N}}F_j\right)^\dagger\left(\frac{1}{\sqrt{M}}E_i\oplus
\frac{1}{\sqrt{N}}F_j\right)\\
&=\sum_{i\in N,j\in M} \frac{1}{M}E_i^\dagger E_i\oplus  \frac{1}{N}F_j^\dagger F_j\\
&=\sum_{i\in N}E_i^\dagger E_i\oplus \sum_{j\in M} F_j^\dagger F_j\\
 &=I_{{\cal H}_1}\oplus I_{{\cal H}_2}.
\end{align}
Hence, for any $\sigma\in\Sigma$, ${\cal E}_\sigma$ constructed
above is a trace-preserving quantum operation acting on ${\cal
H}$.
  Therefore, by letting $P_{acc}=P^{(1)}_{acc}\oplus
P^{(2)}_{acc}$,  we get an MO-1gQFA ${\cal M}=\{{\cal
H},\Sigma,\rho_0, \{{\cal E}_\sigma\}_{\sigma\in\Sigma},
P_{acc}\}.$

Furthermore, for any $\rho=\rho_1\oplus\rho_2\in D({\cal H})$
where $\rho_1,\rho_2$ are density operators up to some
coefficients, we have
\begin{align}
{\cal E}_\sigma(\rho)&=\sum_{i\in N,j\in
M}\left(\frac{1}{\sqrt{M}}E_i\oplus
\frac{1}{\sqrt{N}}F_j\right)(\rho_1\oplus\rho_2)\left(\frac{1}{\sqrt{M}}E_i\oplus
\frac{1}{\sqrt{N}}F_j\right)^\dagger\label{10}\\
&=\sum_{i\in N,j\in M}\frac{1}{M}E_i\rho_1E_i^\dagger\oplus\frac{1}{N} F_j\rho_2F_j^\dagger\\
&=\sum_{i\in N}E_i\rho_1E_i^\dagger\oplus\sum_{j\in M}F_j\rho_2F_j^\dagger\\
&={\cal E}^{(1)}_\sigma (\rho_1)\oplus {\cal E}^{(2)}_\sigma
(\rho_2)\label{13}\\
&\in D({\cal H}).
\end{align}
Then it is not difficult to see that for any $x\in \Sigma^*$, we
have
\begin{align}f_{\cal M}(x)=c_1f_1(x)+c_2f_2(x).\end{align}
Thus, we have proved item (ii)  for $k=2$. It is easy to generalize this proof for the general case $k>2$.

(c) Similarly, we prove  item (iii) for $k=2$. Assume that   $f_i$
is induced by MO-1gQFA ${\cal M}_i=\{
H_i,\Sigma,\rho^{(i)}_0,\{{\cal
E}^{(i)}_\sigma\}_{\sigma\in\Sigma}, P^{(i)}_{acc}\}$ for $i=1,2$,
respectively. Then we construct ${\cal M}=\{{\cal
H},\Sigma,\rho_0, \{{\cal E}_\sigma\}_{\sigma\in\Sigma},
P_{acc}\}$ as follows:
\begin{itemize}
    \item ${\cal H}={\cal H}_1\otimes {\cal H}_2$;
    \item $\rho_0=\rho^{(1)}_0\otimes \rho^{(2)}_0$;
    \item ${\cal E}_\sigma={\cal
E}^{(1)}_\sigma\otimes {\cal E}^{(2)}_\sigma$; more specifically,
if ${\cal E}^{(1)}_\sigma$ and ${\cal E}^{(2)}_\sigma$ have
operator element sets $\{E_i\}$ and $\{F_j\}$, then ${\cal
E}_\sigma$ is constructed such that it has operator element set
$\{E_i\otimes F_j\}$;
    \item $P_{acc}=P^{(1)}_{acc}\otimes P^{(2)}_{acc}$.
\end{itemize}
Then, it is easy to see that for any $\sigma\in\Sigma$ and
$\rho=\rho_1\otimes\rho_2\in D({\cal H})$, we have
\begin{align}{\cal
E}_\sigma(\rho)={\cal E}^{(1)}_\sigma(\rho_1)\otimes{\cal
E}^{(2)}_\sigma(\rho_2).\end{align} Furthermore, for any
$x\in\Sigma^*$, we have $f_{\cal M}(x)=f_1(x)f_2(x)$. Thus, we
have proved item (iii)  for $k=2$, and it is easy to generalize this proof for the general case
$k>2$.

 Therefore, we have completed the proof for Theorem
 \ref{Closure}. \qquad\end{proof}
\subsection{ The computational power of MO-1gQFA}\label{Sec-Power-MO-1gQFA}

In this subsection, we investigate the computational power of
MO-1gQFA. In Ref. \cite{Hir08}, Hirvensalo   showed that MO-1gQFA
can simulate any probabilistic automaton. Thus, the languages
recognized by MO-1gQFA with bounded error should contain the set
of all regular languages. Furthermore, with the most general
operations allowed, MO-1gQFA are expected to be  more powerful rather than  recognizing only regular languages.
However, we will prove that the languages recognized by MO-1gQFA
with bounded error are exactly regular languages, despite the
general operations allowed by this model.

In the following,  we first give the formal definition of an
MO-1gQFA recognizing a language with bounded error.
\begin{Df}
 A language $L$ is said to be recognized by MO-1gQFA ${\cal M}$ with
bounded $\epsilon$ ($\epsilon>0$), if for some $\lambda\in(0,1]$,
$f_{\cal M}(x)\geq\lambda+\epsilon$ holds for any $x\in L$, and
$f_{\cal M}(y)\leq\lambda-\epsilon$ holds for any $y\notin L$.
\label{Df-Language}
\end{Df}

Before we start to  prove the regularity of    languages
recognized by MO-1gQFA, we first recall  some useful concepts and
related results in \cite{NC00}. The trace distance between density
operators $\rho$ and $\sigma$ is
\begin{align}D(\rho,\sigma)=||\rho-\sigma||_{tr}\end{align} where
$||A||_{tr}=\text{Tr}\sqrt{A^\dagger A}$ is the trace norm of
operator $A$. The trace distance between two probability
distributions $\{p_x\}$ and $\{q_x\}$ is
\begin{align}
D(p_x,q_x)=\sum_x|p_x-q_x|.
\end{align}
In the following, we recall two results regarding the trace
distance that will be used later on.
\begin{Lm}[\cite{NC00}] Let $\rho$ and $\sigma$ be two density operators. Then we have $$D({\cal E}(\rho),{\cal E}(\sigma))\leq
D(\rho,\sigma)$$
 for any  trace-preserving quantum
operation ${\cal E}$.\label{Trace-Pro1}
\end{Lm}

\begin{Lm}[\cite{NC00}]  Let $\rho$ and $\sigma$ be two density
operators. Then we have
$$D(\rho,\sigma)=\max_{\{E_m\}}D(p_m,q_m)$$ where
$p_m=\text{Tr}(\rho E_m)$, $q_m=\text{Tr}(\sigma E_m)$ and the
maximization is over all POVMs $\{E_m\}$\label{Trace-Pro2}.
\end{Lm}

For the sake of readability, we also recall the Myhill-Nerode theorem in \cite{HU79} in the following.
\begin{Th}[Myhill-Nerode theorem \cite{HU79}]\label{MN-Th}
The following three statements are equivalent:
\begin{enumerate}
  \item The set $L\subseteq\Sigma^*$ is accepted by some finite automata.
    \item $L$ is the union of some  equivalence classes of a right invariant equivalence relation of finite index.
  \item Let equivalence relation $R_L$ be defined by: $xR_Ly$ if and only if for all $z\in\Sigma^*$, $xz$ is in $L$ exactly when $yz$ is in $L$. Then $R_L$ is of finite index.
\end{enumerate}
\end{Th}
 Now, we present the main result of this subsection.
\begin{Th}
The languages recognized by MO-1gQFA with bounded error are
regular.\label{theorem-Power-MO-1gQFA}
\end{Th}

\begin{proof} Assume that $L$ is  recognized by
 MO-1gQFA ${\cal M}=\{ {\cal H},\Sigma,\rho_0,\{{\cal
E}_\sigma\}_{\sigma\in\Sigma}, P_{acc}\}$ with bounded error
$\epsilon$. We define an equivalence
relation ``$\equiv_L$'' on $x,y\in\Sigma^*$ such that $x\equiv_L y$ if
for any $z\in\Sigma^*$, $xz\in L$ iff $yz\in
L$. Then in terms of Theorem \ref{MN-Th}, if we can prove that  the number of
equivalence classes induced by ``$\equiv_L$'' is finite, then $L$ is regular.

Let $S=\{A: \|A \|_{tr}\leq1, \text{and}~A~\text{is a linear
operator on}~ {\cal H}\}$. Then $S$ is
a bounded subset from a finite-dimensional space.
Let $\rho_x={\cal E}_{x_n}\circ\dots\circ{\cal E}_{x_2}\circ{\cal
E}_{x_1}(\rho_0)$, i.e.,  the state of ${\cal M}$ after having
been fed with word $x$. Then for every $x\in\Sigma^{*}$, it can be
seen that $\rho_x\in S$, since we have
$||\rho_x||_{tr}=\text{Tr}(\rho_x)=\text{Tr}(\rho_0)=1$, where the
second equality holds, because every operation used is
trace-preserving. Now, suppose that $x\not\equiv_L y$, that is,
there exists a string $z\in\Sigma^*$ such that $xz\in L$ and
$yz\notin L$. Then we have
\begin{align}
\text{Tr}(P_{acc}{\cal
E}_z(\rho_x))\geq\lambda+\epsilon~~\text{and}~~\text{Tr}(P_{acc}{\cal
E}_z(\rho_y))\leq\lambda-\epsilon
\end{align}
for some $\lambda\in(0,1]$, where ${\cal E}_z$ stands for ${\cal
E}_{z_n}\circ\dots\circ{\cal E}_{z_2}\circ{\cal E}_{z_1}$.  Denote
$p_{acc}=\text{Tr}(P_{acc}{\cal E}_z(\rho_x))$,
$p_{rej}=\text{Tr}(P_{rej}{\cal E}_z(\rho_x))$,
$q_{acc}=\text{Tr}(P_{acc}{\cal E}_z(\rho_y))$, and
$q_{rej}=\text{Tr}(P_{rej}{\cal E}_z(\rho_y))$. Then in terms of
Lemma \ref{Trace-Pro2}, we have
\begin{align}
||{\cal E}_z(\rho_x)-{\cal
E}_z(\rho_y)||_{tr}\geq|p_{acc}-q_{acc}|+|p_{rej}-q_{rej}|\geq2\epsilon.
\end{align}
On the other hand,  by    Lemma \ref{Trace-Pro1}, we have
\begin{align}
||\rho_x-\rho_y||_{tr}\geq||{\cal E}_z(\rho_x)-{\cal
E}_z(\rho_y)||_{tr}.
\end{align}
Consequently, for any two  strings $x,y\in\Sigma^*$ satisfying
$x\not\equiv_L y$, we always have
\begin{align}
||\rho_x-\rho_y||_{tr}\geq2\epsilon.\label{X-Y}
\end{align}

Now, suppose that $\Sigma^{*}$ consists of infinite equivalence
classes, say $[x^{(1)}],[x^{(2)}]$, $[x^{(3)}],\cdots$. Then by the
boundedness of $S$ from a finite-dimensional space, from the sequence
$\{\rho_{x^{(n)}}\}_{n\in N}$, we can extract
 a Cauchy sequence $\{\rho_{x^{(n_k)}},\}_{k\in N}$, i.e., a  convergent subsequence.
Thus, there exist $x$ and $y$ satisfying $x\not\equiv_L y$ such that
\begin{align}
||\rho_x-\rho_y||_{tr}< 2\epsilon,
\end{align}
which contradicts Ineq. \eqref{X-Y}. Therefore,  the number of the
equivalence classes in $\Sigma^{*}$ induced by the equivalence
relation ``$\equiv_L$'' must
be finite, which implies that $L$ is a regular language.
\qquad\end{proof}

\begin{Rm}
The idea of the above proof is essentially the same as the one in Rabin's seminar paper \cite{Rab63} where it was proved that probabilistic automata  recognize only  regular languages with bounded error. However, some technical treatment is required to adjust it to the case of MO-1gQFA. We also note that from the standpoint of topological space, Jeandel \cite{Jea07}  offered some more general and abstract conditions for the regularity of the languages recognized by an
automaton.
\end{Rm}

\begin{Rm}
It can be seen that the above proof is succinct and without
loss of generality. If we  apply this proof to the special case of
MO-1gQFA---MO-1QFA, then we can prove the regularity of the
languages recognized by MO-1QFA more simply than before. The LQFA introduced by Ambainis {\it et al} \cite{Amb06}  are
also  a special case of MO-1gQFA. Thus, from Theorem
\ref{theorem-Power-MO-1gQFA} it follows straightforward that the
languages recognized by  LQFA with bounded error are
in the class of regular languages. Note that Ambainis {\it et al}
\cite{Amb06} characterized the languages recognized by LQFA using an algebraic approach.
\end{Rm}

The model of MO-1gQFA is  a very generalized  model, which not
only includes these familiar QFA mentioned above, but also can
simulate the classical automata---DFA and even probabilistic
automata. Therefore, we have the following result which is owed to Hirvensalo \cite{Hir08}.
\begin{Th}
MO-1gQFA recognize all regular languages with certainty.
\end{Th}
\begin{proof} The proof is to simulate any probabilistic automaton
by an MO-1gQFA. Indeed, Hirvensalo \cite{Hir08} has already
presented such a simulating process. For the sake of readability, we reproduce
the simulating process in more detail here.

First  recall that an $n$-state probabilistic automaton   ${\cal
A}$ can be represented  as
\begin{align}{\cal A}=(\pi,\Sigma,
\{A(\sigma):\sigma\in\Sigma\},\eta),\label{PA}\end{align}
 where $\pi$ is a stochastic
$n$-dimensional row vector,  $\eta$ is an $n$-dimensional column
vector whose entries are $0$s or $1$s, and for each $\sigma\in
\Sigma$, $A(\sigma)=[A(\sigma)_{ij}]$ is a stochastic $n\times n$
matrix (i.e., each row of it is a stochastic vector), where
$A(\sigma)_{ij}$ means the probability of ${\cal A}$ going to
state $q_j$, given it had been in state $q_i$ and fed with the
symbol $\sigma$. The probability of probabilistic automaton ${\cal
A}$ accepting a string $x_1x_2\dots x_m\in\Sigma^*$ is defined as
\begin{align}
P_{\cal A}(x_1x_2\dots x_m)=\pi A(x_1)A(x_2)\dots A(x_m)\eta.
\end{align}

Now to simulate the above probabilistic automaton ${\cal A}$, we
construct an MO-1gQFA ${\cal M}=\{ {\cal H},\Sigma,\rho_0,\{{\cal
E}_\sigma\}_{\sigma\in\Sigma}, P_{acc}\}$ such that ${\cal
H}=span\{|q_1\rangle,|q_2\rangle,\dots, |q_n\rangle\}$,
$\rho_0=\sum_i\pi_i|q_i\rangle\langle q_i|$, and
$P_{acc}=\sum_{i:\eta_i=1}|q_i\rangle\langle q_i|$. For each
stochastic matrix $A(\sigma)$ in  probabilistic automaton ${\cal
A}$, define a set of operators
$\{E_{ij}=\sqrt{A(\sigma)_{ij}}|q_j\rangle\langle
q_i|:i,j=1,2,\dots, n\}$. Then a direct calculation shows that
$\sum_{i,j=1}^n E_{ij}^\dagger E_{ij}=I$. Thus, a trace-preserving
quantum operation is defined as
\begin{align}
{\cal E}_{\sigma}(\rho)=\sum_{i,j}^n E_{ij}\rho E_{ij}^\dagger.
\end{align}
The action of ${\cal E}_{\sigma}$ on a pure state $|q_i\rangle\langle q_i|$ is as follows:
\begin{align}{\cal E}_{\sigma}(|q_i\rangle\langle q_i|)=\sum_{j=1}^nA(\sigma)_{ij}|q_j\rangle\langle q _j|,\end{align}
which means that  under the operation ${\cal E}_{\sigma}$, state
$|q_i\rangle$ evolves into $|q_j\rangle$ with probability
$A(\sigma)_{ij}$. This is consistent with the action of
$A(\sigma)$ in probabilistic automata ${\cal A}$. Assume that
after having read input $x$, the states of ${\cal A}$ and ${\cal M}$
 are $\pi_x=(\pi_1,\pi_2,\cdots, \pi_n)$ and
$\rho_x=\sum p_i|q_i\rangle\langle q_i|$, respectively. Then by
induction on the length of $x$, it is easy to verify that
$\pi_i=p_i$ holds for $i=1,2,\cdots,n$. Therefore,  MO-1gQFA
${\cal M}$ and probabilistic automaton ${\cal A}$ defined above
have the same accepting probability for each string
$x\in\Sigma^*$.

From  the above process,  a DFA  as a special probabilistic
automaton can be simulated exactly by an MO-1gQFA, and thus, for
every regular language, there is an MO-1gQFA recognizing it with
certainty. \qquad\end{proof}

Furthermore, we can show that the computational power of MO-1gQFA
equals  to that of the QFA defined in \cite{Cia01} and in
\cite{Pas00}. To see that, we first show that the two kinds of QFA
defined in \cite{Cia01, Pas00} are special cases of MO-1gQFA.  We
explain this point in detail for the model in \cite{Pas00}, and it
is similar for the one in \cite{Cia01}.

Unitary transitions are thought to  be a strong restriction on
QFA, and thus limit the computational power of QFA. Then Paschen
\cite{Pas00} proposed a new QFA by adding some ancilla qubits to
avoid the  restriction of unitarity. This is done by adding an
output alphabet. Formally, we have
\begin{Df}[\cite{Pas00}]
An ancilla QFA is a 6-tuple ${\cal M} = (Q,\Sigma, \Omega,\delta,
q_0, F)$, where $Q$ is a finite state set, $\Sigma$ is a finite
input alphabet, $q_0\in Q$ is the initial state,   $F\subseteq Q$
is the set of accepting states, $\Omega$ is an output alphabet,
and the transition function $\delta$ : $Q\times \Sigma\times Q
\times\Omega ~\longrightarrow \mathbb{ C}$ satisfies
\begin{align}
\sum_{p\in Q,
\omega\in\Omega}\delta(q_1,\sigma,p,\omega)^*\delta(q_2,\sigma,p,\omega)=\begin{cases}1,&
q_1=q_2\\
0,&q_1\neq q_2 \end{cases}
\end{align}
for all states $q_1, q_2\in Q$ and $\sigma\in\Sigma$.
\end{Df}

The transition function $\delta$ corresponding to the input symbol
$\sigma\in \Sigma$  can be described by  an isometry   mapping
$V_\sigma$ from $Q$ to $Q\times \Omega$. Suppose the current state
of ${\cal M}$ defined above is $\rho$. Then after  reading
$\sigma$, the state of ${\cal M}$ evolves to
\begin{align}
\rho'=\text{Tr}_\Omega(V_\sigma\rho V_\sigma^\dagger).
\end{align}
Recalling the Stinespring representation (Eq.
\eqref{SR-operation}) of quantum operations, it is easy to see
that the state of ${\cal M}$ evolves by a trace-preserving quantum
operation. Thus, an ancilla QFA is just a special MO-1gQFA.

 Therefore,  the language recognized
by an ancilla QFA with bounded error is a regular language.  On
the other hand, in Ref. \cite{Pas00}, it was proved that ancilla
QFA can recognize any regular language with certainty. Hence, the
languages recognized by ancilla QFA with bounded error are exactly
regular languages.

 Following the idea in Bennett \cite{Ben73},
Ciamarra  \cite{Cia01} proposed a new model of 1QFA
 whose computational power was shown to be  at least equal to
that of classical automata. For convenience,  we call the QFA
defined in \cite{Cia01} as {\it Ciamarra  QFA} named after the
author. Similar to the above process, it is not difficult to see
that the internal state of a Ciamarra  QFA evolves by a
trace-preserving quantum operation, and thus,  a Ciamarra  QFA is
also a special MO-1gQFA.

In summary, we have the following result.
\begin{Co}
The ancilla QFA in \cite{Pas00} and the Ciamarra QFA in
\cite{Cia01} are both special cases of MO-1gQFA. Furthermore, the
three kinds of QFA have the same computational power, and
recognize exactly  regular languages with bounded error.
\end{Co}

\subsection{ The equivalence problem of  MO-1gQFA} \label{Eqv}
In this subsection, we discuss the equivalence problem of
MO-1gQFA. As we know, determining the equivalence between
computing models is of importance   in the theory of classical
computation. For example, determining whether two DFA are
equivalent is an important problem in the theory of classical
automata \cite{HU79}, and determining whether two probabilistic
automata  are equivalent has also been deeply studied
\cite{Paz71,Tze92}. Similarly, the equivalence problem for quantum
computing models is also worth studying, which may redounds to
clarifying the essential difference between quantum and classical
computing models. Indeed, there has already been some work done on
the equivalence problem for quantum automata
\cite{LQ06,LQ08a,LQ08b,QY09,QZL09}.

In the following,  we first give the formal definition of the
equivalence between two MO-1gQFA.
\begin{Df}
Two MO-1gQFA ${\cal M}_1$ and  ${\cal M}_2$ on the same input
alphabet $\Sigma$ are said to be equivalent ($k$-equivalent,
resp.), if $f_{{\cal M}_1}(w)=f_{{\cal M}_2}(w)$ holds for any
$w\in\Sigma^*$ (for any $w\in\Sigma^*$ with $|w|\leq k$, resp.).
\end{Df}

 Given an MO-1gQFA ${\cal M}=\{ {\cal H},\Sigma,\rho_0,\{{\cal
E}_\sigma\}_{\sigma\in\Sigma}, P_{acc}\}$,  denote ${\cal
E}_x={\cal E}_{x_n}\circ\cdots\circ{\cal E}_{x_2}\circ{\cal
E}_{x_1}$ and  $\rho_x={\cal E}_x(\rho_0)$ for $x=x_1x_2\cdots
x_n$. For a vector set $S$, $span S$ denotes the linear space
spanned by $S$. Then we give a key lemma in the following.
\begin{Lm}
 For an MO-1gQFA ${\cal M}=\{
{\cal H},\Sigma,\rho_0,\{{\cal E}_\sigma\}_{\sigma\in\Sigma},
P_{acc}\}$, denote
\begin{align}
\varphi(k)=span\{\rho_x:\rho_x={\cal E}_x(\rho_0), |x|\leq k\}.
\end{align}
Then there exists an integer $i_0\leq n^2$ such that
$\varphi(i_0)=\varphi(i_0+j)$ for $j=1,2,\cdots$, where $n=\dim
{\cal H}$.\label{Lm-Eqv}
\end{Lm}

\begin{proof}  First, from the definition of $\varphi(k)$,
it is readily seen that
\begin{align}
1\leq\dim~\varphi(1)\leq\dim~\varphi(2)\leq\cdots\leq\dim~\varphi(i)\leq\cdots\leq
n^2.
\end{align}
Thus, there exists an integer $i_0\leq n^2$ such that
$\varphi(i_0)=\varphi(i_0+1)$. Next we prove that
$\varphi(i_0)=\varphi(i_0+j)$ holds for $j=2,3,\cdots$. Without
loss of generality, we prove that $\varphi(i_0)=\varphi(i_0+2)$.
Firstly, from the  fact that $\varphi(i_0)=\varphi(i_0+1)$, for
any $\rho\in \varphi(i_0+1)$, we have
\begin{align}
\rho=\sum_i \alpha_i \rho_{x_i},~~~~\forall x_i:|x_i|\leq i_0.
\end{align}
 Then for any $\rho'\in \varphi(i_0+2)$, we have
\begin{align}
\rho'&=\sum_j \beta_j\rho_{x_j} &&|x_j|\leq i_0+2\\
     &=\sum_j \beta_j{\cal E}_{\sigma_j}(\rho_{x'_j})&& |x'_j|\leq
     i_0+1\\
     &=\sum_j \beta_j{\cal E}_{\sigma_j}\left( \sum_i\alpha_i
     \rho_{x''_i}\right) && |x''_j|\leq
     i_0\\
     &=\sum_{i,j}\alpha_i\beta_j {\cal E}_{\sigma_j}(
     \rho_{x''_i})&& |x''_j|\leq
     i_0\\
     &=\sum_{i,j}\alpha_i\beta_j\rho_{x''_{ij}} && |x''_{ij}|\leq
     i_0+1\\
     &\in \varphi(i_0+1).
\end{align}
Hence, we have $\varphi(i_0)=\varphi(i_0+2)$. Similarly, we can
show that $\varphi(i_0)=\varphi(i_0+j)$ for $j\geq 3$. This ends
the proof.\qquad\end{proof}

Note that in the above proof, we used only the linearity  but no
more properties of quantum operations.\label{Rm-eqv-MO}

Based on the above lemma, we have the following theorem.
\begin{Th}
Two MO-1gQFA ${\cal M}_i=\{{\cal H}_i,\Sigma,\rho^{(i)}_0,\{{\cal
E}^{(i)}_\sigma\}_{\sigma\in\Sigma}, P^{(i)}_{acc}\}~ (i=1,2)$ on
the same input alphabet $\Sigma$ are equivalent if and only if
they are $(n_1+n_2)^2$-equivalent, where $n_i=\dim {\cal H}_i$ for
$i=1,2$.\label{Eqv-MO}
\end{Th}

\begin{proof} The necessity is obvious. So we verify the
sufficiency.  For the two MO-1gQFA $${\cal M}_i=\{ {\cal
H}_i,\Sigma,\rho^{(i)}_0,\{{\cal
E}^{(i)}_\sigma\}_{\sigma\in\Sigma}, P^{(i)}_{acc}\}~~ (i=1,2),$$
denote that ${\cal H}={\cal H}_1\oplus {\cal H}_2$,
$\rho_0=\frac{1}{2}\left (\rho^{(1)}_0\oplus \rho^{(2)}_0\right)$,
and ${\cal E}_\sigma={\cal E}^{(1)}_\sigma\oplus {\cal
E}^{(2)}_\sigma$ for any $\sigma\in\Sigma$. More specifically,
similar to the construction process in Section \ref{3.1}, if
${\cal E}^{(1)}_\sigma$ and ${\cal E}^{(2)}_\sigma$ have operator
element sets $\{E_i\}_{i\in N}$ and $\{F_j\}_{j\in M}$,
respectively, then ${\cal E}_\sigma$ is constructed such that it
has operator element set $\{\frac{1}{\sqrt{M}}E_i\oplus
\frac{1}{\sqrt{N}}F_j\}_{i\in N,j\in M}$. Then ${\cal E}_\sigma$
is a trace-preserving quantum operation for any $\sigma\in\Sigma$,
and from Eq.\eqref{10}--\eqref{13}, we have
\begin{align}
{\cal E}_x(\rho_0)=\frac{1}{2}{\cal E}_x^{(1)}(\rho_0^{(1)})\oplus\frac{1}{2}{\cal E}_x^{(2)}(\rho_0^{(2)}).
\end{align}

Let $P=-P^{(1)}_{acc}\oplus P^{(2)}_{acc}$. Then for any
$x\in\Sigma$, we have
\begin{align}
\text{Tr}\left(P{\cal
E}_x(\rho_0)\right)&=\frac{1}{2}\text{Tr}\left(P^{(2)}_{acc}{\cal
E}^{(2)}_x(\rho^{(2)}_0)\right)-\frac{1}{2}\text{Tr}\left(P^{(1)}_{acc}{\cal
E}^{(1)}_x(\rho^{(1)}_0)\right)\\
&=\frac{1}{2}f_{{\cal M}_2}(w)-\frac{1}{2}f_{{\cal M}_1}(w).
\end{align}
Hence, ${\cal M}_1$ and ${\cal M}_2$ are equivalent if and only if
$\text{Tr}\left(P{\cal E}_x(\rho_0)\right)=0$ holds for any
$x\in\Sigma^*$.

Denote
\begin{align}\varphi(k)=span\{\rho_x:\rho_x={\cal E}_x(\rho_0), |x|\leq k\}.\end{align}
Then from Lemma \ref{Lm-Eqv}, it follows that there exists an
integer $i_0\leq(n_1+n_2)^2$ $(n_i=\dim {\cal H}_i$ for $i=1,2$)
such that $ \varphi(i_0)=\varphi(i_0+j)$ holds for $j=1,2,\cdots$.
Thus, for any $|x|>(n_1+n_2)^2$, ${\cal E}_x(\rho_0)$ can be
linearly represented  by some elements in $\{{\cal E}_y(\rho_0):
|y|\leq(n_1+n_2)^2\}$. Therefore, if $\text{Tr}\left(P{\cal
E}_x(\rho_0)\right)=0$ holds for $|x|\leq(n_1+n_2)^2$, then so
does it for any $x\in\Sigma^*$. We have proved this
theorem.\qquad\end{proof}

\begin{Rm} The above result can be seen as  a generalized version of the one about the equivalence problem of {\it quantum sequential machines} given in \cite{LQ06} by some of us. Thus, the result without loss of generality can be applied to more models.
For instance,  MO-1QFA \cite{MC00},   LQFA \cite{Amb06}, ancilla
QFA \cite{Pas00}, and Ciamarra QFA \cite{Cia01} can  all be seen as special
cases of MO-1gQFA.  Thus, the equivalence criterion  given in Theorem \ref{Eqv-MO} also holds
for these models. Note that the equivalence problem about these models mentioned above had not been addressed before the result given here, except the one about MO-1QFA, and for MO-1QFA, the equivalence criterion given here  consists with the one given in \cite{LQ08b}.
\end{Rm}

\section{One-way general quantum finite automata (II): MM-1gQFA}
In this section, we study another kind of general quantum finite
automata, called MM-1gQFA. Similar to the case of MO-1gQFA, each
input symbol of MM-1gQFA also induces a trace-preserving quantum
operation. The difference is that in an MM-1gQFA, a measurement
deciding to accept or reject is performed after a
trace-preserving quantum operation on reading each symbol, but in
an MO-1gQFA,  a similar measurement is allowed only after all the input symbols having been scanned.

It is known that MM-1QFA recognize with bounded error more languages than MO-1QFA, and even more than LQFA \cite{Amb06}, which implies that the times of the
measurement performed affect the computational power of one-way
QFA. In the foregoing section, we have proved that MO-1gQFA can
recognize any regular language with bounded error. Hence, if the number of 
times of measurement also affects the computational power of 1gQFA,
then the model of MM-1gQFA should recognize some non-regular languages
with bounded error.

Our main aim in this section is to characterize  the languages
recognized by MM-1gQFA. Also, we will discuss the equivalence
problem of MM-1gQFA.  To address these problems, in Section
\ref{Sec4.1} we first develop some techniques with which we can
simulate an MM-1gQFA by a relaxed version of MO-1gQFA in which
each symbol induces a linear super-operator instead of a
trace-preserving quantum operation. Based on these techniques
developed by us, in Section \ref{Sec4.2} we will prove that the
 languages recognized by MM-1gQFA with bounded error are
exactly regular languages, which are the same as those recognized
by MO-1gQFA. Therefore, the number of times the measurement is performed has no effect on
the computational power of 1gQFA, which is greatly different from
the conventional case in 1QFA. In Section \ref{Sec4.3}, we will discuss the
equivalence problem of MM-1gQFA. It is obvious that the
equivalence problem for MM-1gQFA is more difficult than the one
for MO-1gQFA and  MM-1QFA. Using the techniques developed in
Section \ref{Sec4.1}, the equivalence problem of MM-1gQFA can be
solved successfully.

\subsection{Preprocessing an MM-1gQFA }\label{Sec4.1}
In this subsection, we first give the definitions related to
MM-1gQFA. Afterward, we develop some techniques to transform an
MM-1gQFA to a relaxed version of MO-1gQFA, which are the base for
the next subsections.

\begin{Df}\label{MM-1gQFA}
An MM-1gQFA ${\cal M}$ is a six-tuple ${\cal M}=\{ {\cal
H},\Sigma,\rho_0,\{{\cal
E}_\sigma\}_{\sigma\in\Sigma\cup\{\cent,\$\}}, {\cal
H}_{acc},{\cal H}_{rej}\}$, where ${\cal H}$ is a
finite-dimensional Hilbert space, $\Sigma$ is a finite input
alphabet, $\cent$ and $\$$ are respectively the left end-marker
and the right end-marker, $\rho_0$, the initial state of ${\cal M}$,
is a density operator on ${\cal H}$, ${\cal E}_\sigma$
corresponding to symbol $\sigma$ is a trace-preserving quantum
operation acting on ${\cal H}$, ${\cal H}_{acc}$ and ${\cal
H}_{rej}$ are the ``accepting" and ``rejecting" subspaces of
${\cal H}$, respectively, and they together with another subspace
${\cal H}_{non}$ span the full space ${\cal H}$, that is, ${\cal
H}={\cal H}_{acc}\oplus{\cal H}_{rej}\oplus{\cal H}_{non}$.
 There is a measurement $\{P_{non},P_{acc},P_{rej}\}$, of which the elements in turn
are the projectors onto subspace ${\cal H}_{non}, {\cal H}_{acc}$,
and ${\cal H}_{rej}$, respectively.
\end{Df}

In the above definition, it is assumed that the initial state
$\rho_0$ is a density operator from the subspace ${\cal H}_{non}$,
and has no common part with the other two subspaces. That is,
$supp(\rho_0)\subseteq {\cal H}_{non}$ and $supp(\rho_0)\cap{\cal
H}_{l}=\emptyset$ for $l\in\{acc, rej\}$. This assumption does not affect
the computational power of MM-1gQFA, since we can produce
arbitrary density operator from $\rho_0$ by adjusting  operation
${\cal E}_{\cent}$. In fact, the similar assumption was also made
in the definition of 2QFA \cite{KW97}. \label{Rm5}

 The input
string of MM-1gQFA ${\cal M}$ has this form: $\cent x\$$ with
$x\in\Sigma^*$ and $\cent,\$$ the left end-maker and the right
end-marker, respectively. The behavior of MM-1gQFA is similar to
that of MM-1QFA. Reading each symbol $\sigma$ in the input string,
the machine has two actions: (i) first ${\cal E}_\sigma$ is
performed such that the current state $\rho$ evolves into ${\cal
E}_\sigma(\rho)$; (ii) the measurement
$\{P_{non},P_{acc},P_{rej}\}$ is performed on the state ${\cal
E}_\sigma(\rho)$.  If the result ``acc" (or ``rej") is observed,
the machine halts in an accepting (or rejecting) state with a
certain probability. Neither, with probability
$\text{Tr}(P_{non}{\cal E}_\sigma(\rho))$ the machine continues to
read the next symbol.

Define ${\cal V}=L(\cal H)\times\mathbb{R}\times \mathbb{R}$.
Elements of ${\cal V}$ will represent the total states of ${\cal
M}$ as follows: a machine described by $(\rho, p_{acc},
p_{rej})\in {\cal V}$ has accepted
with probability $p_{acc}$, rejected with probability $p_{rej}$,
and neither with probability $tr(\rho)$ in which case the current
density operator is $\frac{1}{tr(\rho)}\rho$. The evolution of
${\cal M}$ reading symbol $\sigma\in\Sigma\cup\{\cent, \$\}$ can
be described by an operator   ${\cal T}_\sigma$ on ${\cal V}$ as
follows:
\begin{align}
{\cal T}_\sigma: (\rho,p_{acc},p_{rej})\rightarrow (P_{non}{\cal
E}(\rho)P_{non},\text{Tr}(P_{acc}{\cal
E}(\rho))+p_{acc},\text{Tr}(P_{rej}{\cal E}(\rho))+p_{rej}).
\end{align}

We use $f_{\cal M}(x)$ to denote the probability that MM-1gQFA
${\cal M}$ accepts $x\in\Sigma^*$. Then $f_{\cal M}(x)$
accumulates all the accepting probabilities produced on reading
each symbol in the input string $\cent x\$$.

Obviously, an MM-1QFA \cite{AW02} is a special MM-1gQFA, and the
model, named GQFA, defined by Nayak \cite{Nay99} is
also a special case of MM-1gQFA. Thus, all the results obtained
later for MM-1gQFA also hold for the two models.

 It is easy to see that an MO-1gQFA can be simulated by an
MM-1gQFA. Here we ask a  question in the opposite direction: can
an MM-1gQFA be simulated by an MO-1gQFA? If we relax the
definition of MO-1gQFA, we find the answer  is ``yes". To do that,
we first define a model named {\it Measure-Once Linear Machine}
(MO-LM) as follows.
\begin{Df}
An MO-LM,   represented by  ${\cal M}=\{{\cal H},\Sigma,\rho_0,
\{\Theta_\sigma\}_{\sigma\in\Sigma}, P_{acc}\}$, is similar to an
MO-1gQFA,  where all the elements except $\Theta_\sigma$ are the
same as those in MO-1gQFA, and  $\Theta_\sigma: L({\cal
H})\rightarrow L({\cal H})$ is a linear super-operator, not
necessarily a trace-preserving quantum operation.\label{Df-LM}
\end{Df}

An MO-LM  ${\cal M}$ induces a function $f_{\cal M}:
\Sigma^*\rightarrow \mathbb{C} $ as follows:
\begin{align}
f_{\cal M}(x_1x_2\dots
x_m)=\text{Tr}(\Theta_{x_m}\circ\dots\circ\Theta_{x_2}\circ\Theta_{x_1}(\rho_0)P_{acc})
\end{align}
where  $\Theta_2\circ\Theta_1(\rho)$ stands for
$\Theta_2(\Theta_1(\rho))$.

In the following, we decompose each  trace-preserving quantum
operation in an MM-1gQFA  into three parts, which will be useful
when we construct an MO-LM to simulate an MM-1gQFA.

\begin{Lm}
Given a trace-preserving quantum operation ${\cal E}(\rho)=\sum_m
E_m \rho E_m^\dagger$ acting on the finite-dimensional Hilbert
space ${\cal H}={\cal H}_{non}\oplus{\cal H}_{acc}\oplus{\cal
H}_{rej}$, there is a decomposition
$E_m=E_m^{(non)}+E_m^{(acc)}+E_m^{(rej)}$ for every $E_m$,  such
that for any $l\in\{non,acc,rej\}$, there is
\begin{align}
\sum_m {E_m^{(l)}}^\dagger E_m ^{(l)}=I_l
\end{align}
 where $I_l$ is the identity on subspace $H_l$, and for any
positive operator $\rho_l$ on ${\cal H}$ satisfying
$supp(\rho_l)\subseteq H_{l}$ with $\l\in\{non,acc,rej\}$, there is
\begin{align}
{\cal E}(\rho_l)=\sum_m E_m^{(l)} \rho_l
{E_m^{(l)}}^\dagger.\label{eq-qd}
\end{align}\label{Lm-qd}
\end{Lm}
\begin{proof} Let $\{|n_i\rangle\}$, $\{|a_i\rangle\}$,
and $\{|r_i\rangle\}$ be the  orthonormal bases of ${\cal
H}_{non}$,${\cal H}_{acc}$, and ${\cal H}_{rej}$, respectively.
Then $\{|n_i\rangle\}\cup\{|a_i\rangle\}\cup\{|r_i\rangle\}$ form
an orthonormal base of ${\cal H}$, and for simplicity, we refer
this base as $\{|l\rangle\}$. Then each element $E_m$ in the
operator-sum representation of ${\cal E}$ can be represented in
the outer product form with  base $\{|l\rangle\}$ as
\begin{align}
E_m=\sum_{ll^{'}}e_{ll^{'}}|l\rangle\langle l^{'}|.
\end{align}
More specifically,  $E_m$ can be decomposed into three parts
$E_m=E_m^{(non)}+E_m^{(acc)}+E_m^{(rej)}$ where
\begin{align}
&E_m^{(non)}=\sum e_{l n_i}|l\rangle\langle n_i|,\\
&E_m^{(acc)}=\sum e_{l a_i}|l\rangle\langle a_i|,\\
&E_m^{(rej)}=\sum e_{l r_i}|l\rangle\langle r_i|.
\end{align}
Since ${\cal E}$ is trace-preserving, there is
\begin{align}
&\sum_m E_m^\dagger E_m=I\\
\Leftrightarrow&\sum_m\left(E_m^{(non)}+E_m^{(acc)}+E_m^{(rej)}\right)^\dagger
\left(E_m^{(non)}+E_m^{(acc)}+E_m^{(rej)}\right)=I \\
\Leftrightarrow&\sum_m\left(\sum_l{E_m^{(l)}}^\dagger
E_m^{(l)}+\sum_{l\neq l^{'}}{E_m^{(l)}}^\dagger
E_m^{(l^{'})}\right)=I.
\end{align}

In the above, we should note that for each $l\in\{non,acc,rej\}$,
${E_m^{(l)}}^\dagger E_m^{(l)}$ includes both  diagonal and
non-diagonal elements of $\sum_m E_m^\dagger E_m$,  and
${E_m^{(l)}}^\dagger E_m^{(l^{'})}$ with $l\neq l^{'}$ includes
only non-diagonal elements of that.  Furthermore, it
should be noticed that ${E_m^{(l)}}^\dagger E_m^{(l)}$ with
$l\in\{non,acc,rej\}$ and ${E_m^{(l)}}^\dagger E_m^{(l^{'})}$ with
$l\neq l^{'}$ do not simultaneously have no-zero elements in the
same position. For example, it is easy to see that
${E_m^{(non)}}^\dagger E_m^{(non)}$  and ${E_m^{(non)}}^\dagger
E_m^{(acc)}$ are  in the following forms:
\begin{align}
&{E_m^{(non)}}^\dagger E_m^{(non)}=\sum e'_{ij}|n_i\rangle\langle
n_j|,\\
&{E_m^{(non)}}^\dagger E_m^{(acc)}=\sum e'_{ij}|n_i\rangle\langle
a_j|.
\end{align}
Obviously, they do not simultaneously have   no-zero elements in
the same position. Similarly, we can verify the other cases.

Therefore, from the equality $\sum_m E_m^\dagger E_m=I$ we
conclude that \begin{align}\sum_m\sum_{l\neq
l^{'}}{E_m^{(l)}}^\dagger E_m^{(l^{'})}=0,\end{align} and
\begin{align}
\sum_m {E_m^{(non)}}^\dagger E_m^{(non)}+\sum_m
{E_m^{(acc)}}^\dagger E_m^{(acc)}+\sum_m {E_m^{(rej)}}^\dagger
E_m^{(rej)}=I.
\end{align}
At the same time, we note that
\begin{align}
supp({E_m^{(l)}}^\dagger E_m^{(l)})\subseteq {\cal H}_{l}
\end{align}
holds for each $l\in\{non,acc,rej\}$. Thus we have
\begin{align}
\sum_m {E_m^{(l)}}^\dagger E_m ^{(l)}=I_l
\end{align}
for each $l\in\{non,acc,rej\}$, where $I_l$ is the identity on
subspace ${\cal H}_l$.

Next, we prove Eq. \eqref{eq-qd} for the case $l=non$. Given a
positive operator $\rho_{non}$ satisfying
$supp(\rho_{non})\subseteq
 {\cal H}_{non}$, it is easy to verify that
 \begin{align}
E_m ^{(l)}\rho_{non}=\rho_{non}{E_m ^{(l)}}^{\dagger}=0
~\text{for}~ l\in\{acc,rej\}.
 \end{align}
Thus we have
\begin{align}
{\cal E}(\rho_{non})=\sum_m E_m^{(non)} \rho_{non}
{E_m^{(non)}}^\dagger.
\end{align}

Similarly, we can also prove  Eq. \eqref{eq-qd} for the other
cases where $l\in\{acc,rej\}$. Hence, we have completed the proof
of Lemma \ref{Lm-qd}.\qquad\end{proof}

Now we are in a position to simulate an MM-1gQFA by an MO-LM.
\begin{Th}
An MM-1gQFA ${\cal M}=\{ {\cal H},\Sigma,\rho_0,\{{\cal
E}_\sigma\}_{\sigma\in\Sigma\cup\{\cent,\$\}}, {\cal
H}_{acc},{\cal H}_{rej}\}$ can be simulated by an MO-LM ${\cal
M^{'}}=\{{\cal H},\Sigma,\rho_0,
\{\Theta_\sigma\}_{\sigma\in\Sigma\cup\{\cent,\$\}}, P_{acc}\}$,
such that $f_{\cal M}(x)=f_{\cal M^{'}}(\cent x\$)$ holds for each
$x\in\Sigma^*$.\label{Simulate}
\end{Th}
\begin{proof} Given an  MM-1gQFA ${\cal M}=\{ {\cal
H},\Sigma,\rho_0,\{{\cal
E}_\sigma\}_{\sigma\in\Sigma\cup\{\cent,\$\}}, {\cal
H}_{acc},{\cal H}_{rej}\}$, we construct an MO-LM ${\cal M}^{'}$
such that all the elements except $\Theta$ are the same as those
in  MM-1gQFA ${\cal M}$. Then the key step is to construct a
linear super-operator $\Theta$ to simulate the quantum operation
${\cal E}$ and the measurement $\{P_{non},P_{acc},P_{rej}\}$
performed by ${\cal M}$. We complete this with two steps: (i) first
construct a linear super-operator ${\cal F}$: $L({\cal
H})\rightarrow L({\cal H})$ to simulate the quantum operation
${\cal E}$; (ii) next construct another linear super-operator
${\cal F}^{'}$ to simulate the measurement
$\{P_{non},P_{acc},P_{rej}\}$.

 For the trace-preserving quantum operation in ${\cal M}$: ${\cal
E}(\rho)=\sum_{m=1}^M E_m\rho E_m^\dagger$, in terms of  Lemma
\ref{Lm-qd}, each $E_m$ can be decomposed as
$E_m=E_m^{(non)}+E_m^{(acc)}+E_m^{(rej)}$. Then  we construct a
linear operator on ${\cal H}$ as
 \begin{align}
 F_m=E_m^{(non)}+\frac{1}{\sqrt{M}}P_{acc}+\frac{1}{\sqrt{M}}P_{rej},
 \end{align}
where $M$ is the number of  operators in the operator-sum
representation of ${\cal E}$, and $P_{acc}$ and $P_{rej}$ are the
projectors onto  subspaces ${\cal H}_{acc}$ and ${\cal H}_{rej}$,
respectively. Furthermore, construct a linear super-operator
${\cal F}: L({\cal H})\rightarrow L({\cal H})$ as
\begin{align}
{\cal F}(\rho)=\sum_{m=1}^{M}F_m\rho F_m^\dagger.
\end{align}

Then for any $\rho=\rho_{non}+\rho_{acc}+\rho_{rej}$ satisfying
$supp(\rho_l)\subseteq {\cal H}_l$ with $l\in\{non,acc,rej\}$, we
have
\begin{align}
&{\cal F}(\rho)={\cal F}(\rho_{non})+{\cal F}(\rho_{acc})+{\cal F}(\rho_{rej})\\
=&\sum_{m=1}^{M}E_m^{(non)}{\rho_{non}E_m^{(non)}}^\dagger+\frac{1}{M}\sum_{m=1}^M
P_{acc}\rho_{acc}P_{acc}+
\frac{1}{M}\sum_{m=1}^M P_{rej}\rho_{rej}P_{rej}\\
=&{\cal E}(\rho_{non})+\rho_{acc}+\rho_{rej}.\label{F}
\end{align}
In the above process, we used Lemma \ref{Lm-qd} and such
properties:
\begin{align}
&P_{l}\rho_{non}=0, \rho_{non}P_{l}=0 ~\text{for}~ l\in\{acc, rej\},\\
&E_m^{(non)}\rho_l=0,\rho_l{E_m^{(non)}}^\dagger=0 ~\text{for}~
l\in\{acc, rej\}
~\text{and any}~m, \\
&P_{l}\rho_lP_{l}=\rho_l ~\text{for}~ l\in\{acc, rej\}.
\end{align}

 The next step is to simulate the measurement $\{P_{non},P_{acc},P_{rej}\}$ performed by  MM-1gQFA ${\cal M}$.
To do this, construct a trace-preserving quantum operation ${\cal
F}^{'}$ as follows:
\begin{align}
{\cal F}^{'}(\rho)=P_{non}\rho P_{non}+P_{acc}\rho
P_{acc}+P_{rej}\rho P_{rej}.
\end{align}
For ${\cal F}(\rho)$ given in Eq. \eqref{F}, we have
\begin{align}
{\cal F}^{'}({\cal F}(\rho))&={\cal F}^{'}({\cal E}(\rho_{non}))+{\cal F}^{'}(\rho_{acc})+{\cal F}^{'}(\rho_{rej})\\
&={\cal F}^{'}({\cal E}(\rho_{non}))+\rho_{acc}+\rho_{rej}\\
&=P_{non}{\cal E}(\rho_{non})P_{non}+(\rho_{acc}+P_{acc}{\cal
E}(\rho_{non})P_{acc})+(\rho_{rej}+P_{rej}{\cal
E}(\rho_{non})P_{rej}).
\end{align}

In summary, corresponding to the quantum operation ${\cal E}$ and
 the measurement $\{ P_{non}, P_{acc}, P_{rej}\}$ performed by
MM-1gQFA ${\cal M}$ when reading a symbol, we construct a linear
super-operator $\Theta: L({\cal H})\rightarrow L({\cal H})$  for
MO-LM ${\cal M}^{'}$ by letting $\Theta={\cal F}^{'}\circ{\cal
F}$. Then for any $\rho=\rho_{non}+\rho_{acc}+\rho_{rej}$
satisfying $supp(\rho_l)\subseteq{\cal H}_l$ with
$l\in\{non,acc,rej\}$, we have
\begin{align}
\Theta:\rho_{non}+\rho_{acc}+\rho_{rej}\rightarrow
\rho_{non}^{'}+(\rho_{acc}+\rho_{acc}^{'})+(\rho_{rej}+\rho_{rej}^{'})\label{Theta}
\end{align}
such that $supp(\rho_l^{'})\subseteq{\cal H}_l$ for
$l\in\{non,acc,rej\}$, and more specifically
$\rho_{non}^{'}=P_{non}{\cal E}(\rho_{non})P_{non}$,
$\rho_{acc}^{'}=P_{acc}{\cal E}(\rho_{non})P_{acc}$, and
$\rho_{rej}^{'}=P_{rej}{\cal E}(\rho_{non})P_{rej}$.

Next we should prove that ${\cal M}$ and ${\cal M}^{'}$ have the
same accepting probability for each input string.
 First we mention that the state $\bar{\rho}\in L({\cal H})$ of
 MO-LM ${\cal M}^{'}$  after having read some input string can always be written  in this form:
\begin{align}\bar{\rho}=\rho_{non}+\rho_{acc}+\rho_{rej}\end{align} with
$supp(\rho_l)\subseteq{\cal H}_l$ for $l\in\{non,acc,rej\}$. To
see that, first we note that  the initial state $\rho_0$ is
trivially in the form, and from Eq. \eqref{Theta}, we see that the
linear super-operator $\Theta$ maps a state in the form to another
state in the same form.

Also recall that the state of MM-1gQFA ${\cal M}$ can be described
by an element in ${\cal V}=L(\cal H)\times\mathbb{R}\times
\mathbb{R}$ as
 \begin{align}(\rho, p_{acc},p_{rej}).\end{align}

To prove that ${\cal M}$ and ${\cal M}^{'}$ have the same
accepting probability for each input string, we prove the
following proposition.
\begin{Pp} After reading any string, the state $(\rho, p_{acc},p_{rej})$ of MM-1gQFA ${\cal
M}$ and the state of MO-LM ${\cal M}^{'}$
$\bar{\rho}=\rho_{non}+\rho_{acc}+\rho_{rej}$ where
$supp(\rho_l)\subseteq{\cal H}_l$ for $l\in\{non,acc,rej\}$
 satisfy the following equalities:
\begin{align}
&\rho=\rho_{non},\label{Equality1}\\
&p_{acc}=\text{Tr}(\bar{\rho} P_{acc}).\label{Equality2}
\end{align}\label{Prop}
\end{Pp}
\begin{proof} We prove this proposition by induction on
the length of input string $y$.

{\it Base:} When $|y|=0$, the result holds trivial if only we note
that $supp(\rho_0)\subseteq {\cal H}_{non}$. When $|y|=1$, the
state of ${\cal M}$ evolves as
\begin{align} {\cal T}_y: (\rho_0,0,0)\rightarrow (P_{non}{\cal
E}_y(\rho_0)P_{non},\text{Tr}(P_{acc}{\cal
E}_y(\rho_0)),\text{Tr}(P_{rej}{\cal E}_y(\rho_0))),
\end{align}
and the state of ${\cal M}^{'}$ evolves as
\begin{align}
\Theta_y: \rho_0\rightarrow \bar{\rho}=P_{non}{\cal
E}_y(\rho_0)P_{non}+P_{acc}{\cal E}_y(\rho_0)P_{acc}+P_{rej}{\cal
E}_y(\rho_0)P_{rej}.
\end{align}
Then it is readily seen that Eqs. \eqref{Equality1} and
\eqref{Equality2} hold.

{\it Induction:} Assume that  after having read $y$ with $|y|=k$,
the states of ${\cal M}$ and ${\cal M}^{'}$ are $(\rho,
p_{acc},p_{rej})$ and
$\bar{\rho}=\rho_{non}+\rho_{acc}+\rho_{rej}$, respectively, and
they satisfy
 $\rho=\rho_{non}$ and $p_{acc}=\text{Tr}(\bar{\rho
}P_{acc})$. For $|y|=k+1$, let $y=y^{'}\sigma$ satisfying
$|y^{'}|=k$ and $\sigma\in\Sigma\cup\{\cent, \$\}$. Then the state
of ${\cal M}$ evolves as:
\begin{align}
{\cal T}_\sigma: (\rho,p_{acc},p_{rej})\rightarrow (\rho^{'},
p_{acc}^{'}, p_{rej}^{'}),
\end{align}
where $\rho^{'}=P_{non}{\cal E}_\sigma(\rho)P_{non}$,
$p_{acc}^{'}=\text{Tr}(P_{acc}{\cal E}_\sigma(\rho))+p_{acc}$, and
$p_{rej}^{'}=\text{Tr}(P_{rej}{\cal E}_\sigma(\rho))+p_{rej}$. The
state of ${\cal M}^{'}$  evolves as:
\begin{align}
\Theta_\sigma:
\bar{\rho}=\rho_{non}+\rho_{acc}+\rho_{rej}\rightarrow
\bar{\rho}'=\rho_{non}^{'}+(\rho_{acc}+\rho_{acc}^{'})+(\rho_{rej}+\rho_{rej}^{'}),
\end{align}
where $\rho_{non}^{'}=P_{non}{\cal E}_\sigma(\rho_{non})P_{non}$,
$\rho_{acc}^{'}=P_{acc}{\cal E}_\sigma(\rho_{non})P_{acc}$, and
$\rho_{rej}^{'}=P_{rej}{\cal E}_\sigma(\rho_{non})P_{rej}$.

Then from the assumption $\rho=\rho_{non}$, it is easily seen that
$\rho_{non}^{'}=\rho^{'}$, i.e., Eq. \eqref{Equality1} holds. Also,
we have
\begin{align}
\text{Tr}(\bar{\rho}' P_{acc})&=\text{Tr}((\rho_{acc}+\rho_{acc}^{'}) P_{acc})\\
&=\text{Tr}(\rho_{acc}P_{acc})+\text{Tr}(\rho_{acc}^{'}P_{acc})\\
&=\text{Tr}(\bar{\rho}P_{acc})+\text{Tr}(P_{acc}{\cal E}_\sigma(\rho_{non}))\\
&=p_{acc}+\text{Tr}(P_{acc}{\cal E}_\sigma(\rho))~~~~(\text{by the assumption})\\
&=p_{acc}^{'}.
\end{align}
Thus, we have completed the proof of Proposition
\ref{Prop}.\qquad\end{proof}

From Proposition \ref{Prop}, we know that MM-1gQFA ${\cal M}$ and
MO-LM ${\cal M}^{'}$ have the same accepting probability for any
input string. Therefore, we have completed the proof of Theorem
\ref{Simulate}.\qquad\end{proof}

\begin{Rm}
In the above proof, we should observe the following two points,
which will be useful in the proof of the regularity of languages
recognized by MM-1gQFA in the next subsection:
\begin{enumerate}
\item [(i)] The linear super-operator ${\cal F}: L({\cal
H})\rightarrow L({\cal H})$ defined as ${\cal
F}(\rho)=\sum_{m=1}^{M}F_m\rho F_m^\dagger$ is generally not a
trace-preserving quantum operation, since direct calculation shows
that
\begin{align}
\sum_{m=1}^{M}F_m^\dagger F_m =&\sum_{m=1}^{M}
{E_m^{(non)}}^\dagger E_m^{(non)}+P_{acc}+P_{rej}\\
&~+\frac{1}{\sqrt{M}}\sum_{m=1}^M\left( {E_m^{(non)}}^\dagger
P_{acc}+{E_m^{(non)}}^\dagger
P_{rej}+P_{acc}E_m^{(non)}+P_{rej}E_m^{(non)}\right)
\end{align}
where $\sum_{m=1}^{M} {E_m^{(non)}}^\dagger
E_m^{(non)}+P_{acc}+P_{rej}=I_{\cal H}$. However, it is easy to
see that  for any $\rho=\rho_{non}+\rho_{acc}+\rho_{rej}$
satisfying $supp(\rho_l)\subseteq{\cal H}_l$ with
$l\in\{non,acc,rej\}$, ${\cal F}$ is trace-preserving, i.e.,
$\text{Tr}({\cal F}(\rho))=\text{Tr}(\rho)$.

\item [(ii)] It is not difficult to check that the states of MO-LM
${\cal M}^{'}$ constructed in the above proof are always positive
operators, and for a positive operator $\rho$, there is
$\|\rho\|_{tr}=\text{Tr}(\rho)$.\label{Rm-F}
\end{enumerate}
\end{Rm}
\subsection{The computational power of MM-1gQFA}\label{Sec4.2}
In this subsection, we are going to investigate the language
recognition power of MM-1gQFA. Indeed, characterizing the
languages recognized by various QFA is a central problem in the
study of QFA. At the same time,  making effort to enhance the
computational power of QFA through various strategies (for
example, modifying the definition of QFA) is also an important
issue considered in much work on QFA.

There has been some important work devoted to the characterization
of the language recognition power of MM-1QFA and MO-1QFA.  It is
known that MM-1QFA can recognize more languages with bounded error
than MO-1QFA. For example, MM-1QFA can recognize the language
$a^*b^*$ with bounded error \cite{AF98}, but MO-1QFA can not. From
this fact, we tend to believe that the number of times of the measurement
performed in the computation affects  the computational power of
QFA. Encouraged by this belief, we  have defined the model of MM-1gQFA,
a measure-many version of MO-1gQFA, with hope to enhance the
computational power of 1gQFA. However, in this subsection we will
prove that the languages recognized by MM-1gQFA with bounded error
are exactly regular languages. Thus, MM-1gQFA and MO-1gQFA have
the same computational power.

We first recall some notions and results that
will be used later. In the following theorem, $rank(A)$ denotes the rank of $A$.
\begin{Th}[Singular-Value Theorem \cite{HJ86,NC00}] Let $A: {\cal H}_1\rightarrow
{\cal H}_2$ be a linear operator and let $rank(A)=r$. Then there
exist some positive real numbers $s_1,s_2,\dots,s_r$ and
orthonormal sets
$\{|v_1\rangle,|v_2\rangle,\dots,|v_r\rangle\}\subset {\cal H}_1$
and $\{|u_1\rangle,|u_2\rangle,\dots,|u_r\rangle\}\subset {\cal
H}_2$ such that
\begin{align}
A=\sum_{i=1}^rs_i|u_i\rangle\langle v_i|.
\end{align}
\end{Th}

 We can characterize several important
norms of linear operators using their singular values. There are
different norms for linear operators, and here we present two
 usually used norms: the Frobenius
norm and the trace norm.

The Frobenius norm of $A\in L({\cal H})$ is defined as
\begin{align}||A||_F=\sqrt{\langle A,A\rangle}\end{align} where $\langle
A,B\rangle=\text{Tr}(A^\dagger B)$ is the Hilbert-Schmidt inner
product between $A$ and $B$. Then  the Cauchy-Schwarz inequality
implies
\begin{align}|\langle A,B\rangle|\leq||A||_F||B||_F.\end{align}
Equivalently, the Frobenius norm  $||A||_F$ can be characterized
by  the singular values of $A$  as follows:
\begin{align}
||A||_F=\left(\sum s_i^2\right)^{\frac{1}{2}}.
\end{align}

The  trace norm of   $A\in L({\cal H})$, defined as
$||A||_{tr}=\text{Tr}\sqrt{A^\dagger A}$, will  often be used in the
foregoing sections. Note that if $A$ is a positive operator, then
$||A||_{tr}=\text{Tr}(A)$. Similar to the Frobenius norm, the
trace norm can also be characterized by
 singular values as \begin{align}||A||_{tr}=\sum_i
s_i.\end{align} In terms of the singular values of $A\in L({\cal
H})$, it is not difficult to see
\begin{align}
||A||_F\leq ||A||_{tr}.\label{Ineq-norm}
\end{align}

In fact, different norms defined for $A\in L({\cal H})$ are
equivalent in the following sense.
\begin{Lm} [\cite{HJ86}]
Let $\|\cdot\|_\alpha$ and $\|\cdot\|_\beta$ be any two norms on a
finite dimensional vector space $V$. Then there exist two finite
positive constants $c_1$ and $c_2$ such that
$c_1\|x\|_\alpha\leq\|x\|_\beta\leq c_2\|x\|_\alpha$ for all $x\in
V$.\label{norm-eqv2}
\end{Lm}

Obviously, $L({\cal H})$ is a finite dimensional vector space
given ${\cal H}$ is finite. Thus, the norms on $L({\cal H})$
defined above  satisfy the property given in the above lemma.

  Now for the MO-LM ${\cal M}^{'}=\{{\cal
H},\Sigma,\rho_0, \{\Theta_\sigma\}_{\sigma\in\Sigma\cup\{\cent,
\$\}}, P_{acc}\}$ which was constructed to simulate an MM-1gQFA,
let ${\cal S}=span\{\Theta_{\cent w}(\rho_0):w\in\Sigma^{*}\}$,
where $\Theta_{\sigma_1\sigma_2}$ stands for
$\Theta_{\sigma_2}\circ \Theta_{\sigma_1}$. Then we have the
following result.
\begin{Lm}
 There exists a
constant $c$ such that $\left\|\Theta_{y\$}(\rho)\right\|_{tr}\leq
c\|\rho\|_{tr}$ for any $\rho\in{\cal S}$ and  $y\in\Sigma^{*}$.
\label{Norm-Bound}\end{Lm}

\begin{proof} Firstly find a base for ${\cal S}$ as:
$\rho_1=\Theta_{\cent w_1}(\rho_0),\rho_2=\Theta_{\cent
w_2}(\rho_0),\dots,\rho_m=\Theta_{\cent w_m}(\rho_0)$. For each
$1\leq i\leq m$, let $e_i\in L({\cal H})$ satisfy $||e_i||_F=1$,
$e_i\perp\{\rho_j:j\neq i\}$ and $e_i\not\perp \rho_i$. Note that
for $A,B\in L({\cal H})$, $A\perp B$ means $\langle
A,B\rangle=tr(A^\dagger B)=0$. Then  $\rho\in{\cal S}$ can be
linearly represented as $\rho=\sum_{i=1}^m \alpha_i \rho_i$, and
it holds that
\begin{align}
||\rho||_F\geq|\langle e_i,\rho\rangle|=|\alpha_i|\cdot|\langle
e_i, \rho_i\rangle|.\label{S}\end{align} Hence, we have
\begin{align*}
\|\Theta_{y\$}(\rho)\|_F&=\left\|\sum_{i=1}^m \alpha_i
\Theta_{y\$}(\rho_i)\right\|_F
=\left\|\sum_{i=1}^m \alpha_i \Theta_{\cent w_ix\$}(\rho_0)\right\|_F\\
&\leq\sum_{i=1}^m|\alpha_i|\cdot\|\Theta_{\cent
w_ix\$}(\rho_0)\|_F
\leq\sum_{i=1}^m|\alpha_i|\cdot \left\|\Theta_{\cent w_i x\$}(\rho_0)\right\|_{tr}&&(\text{by Ineq. \eqref{Ineq-norm}})\\
&=\sum_{i=1}^m|\alpha_i|\text{Tr}(\rho_0) =\sum_{i=1}^m|\alpha_i|&&  \\
&\leq||\rho||_F\sum_{i=1}^m1/|\langle e_i,\rho_i\rangle| &&(\text{by Ineq.~\eqref{S}})\\
&= K||\rho||_F,
\end{align*}
where $K=\sum_{i=1}^m 1/|\langle e_i,\rho_i\rangle|$ is a constant
without dependence on $\rho$, and the third equality follows from
the observations (i) and (ii) made at the end of Section
\ref{Sec4.1}. Furthermore, by Lemma \ref{norm-eqv2} and Ineq.
\eqref{Ineq-norm}, we have
\begin{align}
\|\Theta_{y\$}(\rho)\|_{tr}\leq c_1\|\Theta_{y\$}(\rho)\|_F\leq
c_1K||\rho||_F \leq c_1K||\rho||_{tr}.
\end{align}
Thus, by letting $c=c_1K$, we have completed the proof of Lemma
\ref{Norm-Bound}.\qquad\end{proof}

The definition of MM-1gQFA recognizing a language with bounded
error is similar to the one for MO-1gQFA given in Definition
\ref{Df-Language}. In the following, we present a complete
characterization of the languages recognized by MM-1gQFA with
bounded error. We mention that the framework of the following
proof is
 similar to the one in Theorem \ref {theorem-Power-MO-1gQFA}, but there
 needs some new technical treatment.
\begin{Th}
The languages recognized by MM-1gQFA with bounded error are
regular.\label{GQFA}
\end{Th}

\begin{proof}  Assume that $L$ is  recognized by MM-1gQFA ${\cal
M}=\{ {\cal H},\Sigma,\rho_0,\{{\cal
E}_\sigma\}_{\sigma\in\Sigma\cup\{\cent,\$\}}, {\cal
H}_{acc},{\cal H}_{rej}\}$ with bounded error $\epsilon$. Then in
terms of Theorem \ref{Simulate}, there exists an MO-LM  ${\cal
M^{'}}=\{{\cal H},\Sigma,\rho_0,
\{\Theta_\sigma\}_{\sigma\in\Sigma\cup\{\cent,\$\}}, P_{acc}\}$
such that for some $\lambda\in (0,1]$, $f_{{\cal M}^{'}}(\cent
x\$)\geq \lambda+\epsilon$ holds for any $x\in L$, and $f_{{\cal
M}^{'}}(\cent y\$)\leq \lambda-\epsilon$ holds for any $y\notin
L$.

 We define an equivalence
relation ``$\equiv_L$'' on $x,y\in\Sigma^*$ such that $x\equiv_L y$ if
for any $z\in\Sigma^*$, $xz\in L$ iff $yz\in L$. Then in terms of
the Myhill-Nerode theorem (Theorem \ref{MN-Th}), it is sufficient to prove
that the number of equivalence classes induced by ``$\equiv_L$'' is
finite.

Let $S=\{A: \|A \|_{tr}\leq1, \text{and}~A~\text{is a linear
operator on}~ {\cal H}\}$. Then $S$
is a bounded  subset from a finite-dimensional space.
Let
$\rho_x=\Theta_{x_n}\circ\dots\circ\Theta_{x_2}\circ\Theta_{x_1}\circ\Theta_{\cent}(\rho_0)$,
i.e.,  the state of ${\cal M}^{'}$ after having been fed with
input string $\cent x$ with $x\in \Sigma^*$. Then for every $x$,
it can be seen that $\rho_x\in S$, since we have
$||\rho_x||_{tr}={\text Tr}(\rho_x)={\text Tr}(\rho_0)=1$ which
follows from  the observations (i) and (ii) made at the end of
Section \ref{Sec4.1}.  Now, suppose that $x\not\equiv_L y$, that is,
there exists a string $z\in\Sigma^*$ such that $xz\in L$ and
$yz\notin L$. Then we have
\begin{align}
\text{Tr}(P_{acc}\Theta_{z\$}(\rho_x))\geq\lambda+\epsilon~~\text{and}~~\text{Tr}(P_{acc}\Theta_{z\$}(\rho_y))\leq\lambda-\epsilon
\end{align}
for some $\lambda\in(0,1]$, where $\Theta_z$ stands for
$\Theta_{z_n}\circ\dots\circ\Theta_{z_2}\circ\Theta_{z_1}$.

Denote $\overline{P}_{acc}=I-P_{acc}$. Then
$\{P_{acc},\overline{P}_{acc}\}$ is a POVM measurement (exactly
speaking, a projective measurement) on space ${\cal H}$. Note that
Lemma \ref{Trace-Pro2} also holds for any two positive operators.
That is, for any two positive operators $A,B$, it holds that
\begin{align}
||A-B||_{tr}=\max_{\{E_m\}}\sum_{m}|\text{Tr}(E_mA)-\text{Tr}(E_mB)|,
\end{align}
where the maximization is over all POVMs $\{E_m\}$. Indeed, this
property has already been observed in Qiu \cite{Qiu08b}.
Therefore, we have
\begin{align*}
||{\cal E}_{z\$}(\rho_x)-{\cal E}_{z\$}(\rho_y)||_{tr}
&\geq\left|\text{Tr}(P_{acc}{\cal
E}_{z\$}(\rho_x))-\text{Tr}(P_{acc}{\cal
E}_{z\$}(\rho_y))\right|\\
&~~+\left|\text{Tr}(\overline{P}_{acc}{\cal
E}_{z\$}(\rho_x))-\text{Tr}(\overline{P}_{acc}{\cal
E}_{z\$}(\rho_y))\right|\\
 &\geq2\epsilon.
\end{align*}
On the other hand,  in terms of   Lemma \ref{Norm-Bound}, we have
\begin{align}
||\rho_x-\rho_y||_{tr}\geq\frac{1}{c}||{\cal
E}_{z\$}(\rho_x)-{\cal E}_{z\$}(\rho_y)||_{tr},
\end{align}
where $c$ is a constant.
 Consequently, for any two  strings
$x,y\in\Sigma^*$ satisfying $x\not\equiv_L y$, we always have
\begin{align}
||\rho_x-\rho_y||_{tr}\geq\frac{1}{c}2\epsilon.\label{X-Y2}
\end{align}

Now, suppose that $\Sigma^{*}$ consists of infinite equivalence
classes, say $[x^{(1)}]$, $[x^{(2)}]$, $[x^{(3)}],\cdots$. Then by the
boundedness of $S$ from a finite-dimensional space, from the sequence
$\{\rho_{x^{(n)}}\}_{n\in N}$, we can extract
 a Cauchy sequence $\{\rho_{x^{(n_k)}},\}_{k\in N}$, i.e., a  convergent subsequence.
Thus, there exist $x$ and $y$ satisfying $x\not\equiv_L y$ such that
\begin{align}
||\rho_x-\rho_y||_{tr}<\frac{1}{c} 2\epsilon,
\end{align}
which contradicts Ineq. \eqref{X-Y2}. Therefore,  the number of
the equivalence classes in $\Sigma^{*}$ induced by the equivalence
relation ``$\equiv_L$'' must be finite, which implies
that $L$ is a regular language. \qquad\end{proof}

Now we have proved the languages recognized by MM-1gQFA with
bounded error are in the set of regular languages. On the other
hand, it is easy to see that an MO-1gQFA can be simulated by an
MM-1gQFA. Hereby, MM-1QFA can recognize any regular language with
bounded error. Therefore, we have the following result.
\begin{Th}
The languages recognized by MM-1gQFA with bounded error are
exactly regular languages.
\end{Th}

 \begin{Rm}As we know, so far no QFA with a one-way tape head can recognize
a language out of the scope of regular languages. Although we
allow the most general operations---trace-preserving quantum
operations,  QFA with one-way tape heads  still recognize only
regular languages. On the other hand, the two-way QFA defined in
\cite{KW97} can recognize some non-regular languages. Thus, the
uppermost factor affecting the computational power of a QFA should
be the moving direction of its tape head, but not the operations
induced by the input alphabet.
\end{Rm}

\subsection{The equivalence problem of MM-1gQFA}\label{Sec4.3}
In this subsection, we discuss the equivalence problem of
MM-1gQFA. In Section \ref{Eqv}, we have dealt with the equivalence
problem of MO-1gQFA. Apparently, the equivalence problem of
MM-1gQFA is more difficult than that of MO-1gQFA. However, based
on the techniques developed in Section \ref{Sec4.1}, the
equivalence problem of MM-1gQFA can be proved in the same way we did
for MO-1gQFA.

 The formal definitions related  to the equivalence of MM-1gQFA
 are similar to those for MO-1gQFA given in Section \ref{Eqv},
 and  we do not repeat them here. Our result is as follows.
\begin{Th}
Two MM-1gQFA ${\cal M}_i=\{ {\cal H}_i,\Sigma,\rho_0^{(i)},\{{\cal
E}_\sigma^{(i)}\}_{\sigma\in\Sigma\cup\{\cent,\$\}}, {\cal
H}_{acc}^{(i)},{\cal H}_{rej}^{(i)}\}$ with $i=1,2$ are equivalent
if and only if they are $(n_1+n_2)^2$-equivalent, where
$n_i=dim({\cal H}_i)$ for $i=1,2$.\label{Eqv-MM}
\end{Th}
\begin{proof} In terms of Theorem \ref{Simulate}, we know that  two
MM-1gQFA ${\cal M}_i$ with $i=1,2$ can be simulated by two MO-LM
${\cal M}_i^{'}=\{{\cal H}_i,\Sigma,\rho_0^{(i)},
\{\Theta_\sigma^{(i)}\}_{\sigma\in\Sigma\cup\{\cent,\$\}},
P_{acc}^{(i)}\}$, respectively. Then we need only to determine the
equivalence between ${\cal M}_1^{'}$ and ${\cal M}_2^{'}$.

Note that the only difference between MO-1gQFA and MO-LM is
 that ${\cal E}_\sigma$ is a trace-preserving quantum operation
 while $\Theta_\sigma$ is a general linear super-operator. Also, note that  we used only the linearity
 but no more properties of  ${\cal E}_\sigma$ in the
 proof of Lemma \ref{Lm-Eqv}. Therefore, Lemma \ref{Lm-Eqv} also holds
  for MO-LM. Furthermore, using the similar techniques used in
 the proof of Theorem \ref{Eqv-MO}, we obtain the result stated
 in the above theorem. \qquad\end{proof}

\begin{Rm}
The above result also holds for the two special cases of
MM-1gQFA: MM-1QFA \cite{AW02} and GQFA \cite{ANT02}.  Note that the equivalence problem of
MM-1QFA has already been considered in \cite{LQ08a}. In the above,  viewing MM-1QFA as a special case of MM-1QFA, we have  obtained an equivalence criterion slightly different from the one in \cite{LQ08a}. In fact, here we have used a method different from the one in \cite{LQ08a}.
 The equivalence problem of GQFA had not been discussed
before the above result, and here we have addressed this problem.
\end{Rm}

\section{Conclusion}
In this paper, we have studied the model of one-way general
quantum finite automata (1gQFA), in which each symbol in the input
alphabet induces a trace-preserving quantum operation, instead of
a unitary transformation. We have studied two typical models of
1gQFA: MO-1gQFA where a measurement deciding to accept or reject
is allowed only at the end of a computation, and  MM-1gQFA where a
similar measurement is allowed at reading each symbol during a
computation.

We have proved that the languages recognized by MO-1gQFA with
bounded error are still in the scope of regular languages, despite
the most general operations allowed by this model. More exactly,
MO-1gQFA recognize exactly regular languages with bounded error.  Also, two types of QFA
defined in \cite{Cia01, Pas00} which were expected to be more
powerful than MO-1QFA, have been shown to be special cases of MO-1gQFA, and have
the same computational power as MO-1gQFA. We have discussed the
equivalence problem of MO-1gQFA, and it has been proved that two
MO-1gQFA ${\cal M}_1$ and ${\cal M}_2$ are equivalent if and only
they are $(n_1+n_2)^2$-equivalent, where $n_1$ and $n_2$ are the
dimensions of the Hilbert spaces that ${\cal M}_1$ and ${\cal
M}_2$ act on, respectively. In addition, some closure properties
of MO-1gQFA have been presented.

The number of times the measurement is performed is generally thought to affect
the computational power of 1QFA. With this belief, we have defined
the model of MM-1gQFA, a measure-many version of MO-1gQFA.
However, we have proved that MM-1gQFA recognize with bounded error
the same class of languages as MO-1gQFA. Hence, the measurement times
  have no effect on the computational power of
1gQFA, which is greatly different from the conventional case where
MM-1QFA recognize more languages than MO-1QFA \cite{BP99}. Also,
we have addressed the equivalence problem of MM-1gQFA. We have
proved that the equivalence criterion for MO-1gQFA given above
also holds for MM-1gQFA.  The solution of all the above problems
regarding MM-1gQFA is based on such a result proved by us that an
MM-1gQFA can be simulated by a relaxed version of
MO-1gQFA---MO-LM, in which each symbol in the input alphabet
induces a general linear super-operator, not necessarily a
trace-preserving quantum operation.

 From the study in this paper, we have seen that so far no quantum
 finite automaton with a one-way  tape head can recognize with bounded
 error a language out of the scope of regular languages, even if
 the most general operations---trace-preserving quantum operations
 are allowed. On the other hand, we recall that 2QFA introduced by
Kondacs and Watrous \cite{KW97} can recognize the non-regular
language $L_{eq}=\{a^{n}b^{n}|n>0\}$ in linear time. Therefore, it
may be asserted that the uppermost factor affecting the
computational power of QFA is the moving direction of the tape
head,  neither the operation induced by the input symbol, nor the number of  times the measurement is performed.

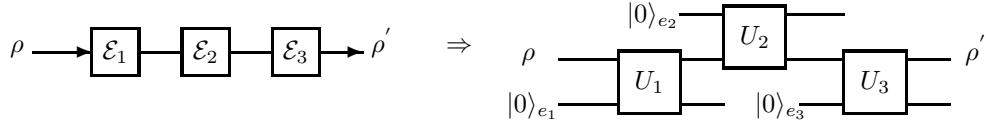
\begin{figure}[htbp]%
\setlength{\unitlength}{1cm}
\begin{picture}(15,4.0)\thicklines
\put(0.2,3){\text{$\rho$}}
 \put(0.5,3){\vector(1,0){0.8}}
 \put(1.3,2.7){\framebox(0.6,0.6){${\cal E}_1$}}
  \put(1.9,3){\line(1,0){0.6}}
 \put(2.5,2.7){\framebox(0.6,0.6){${\cal E}_2$}}
  \put(3.1,3){\line(1,0){0.6}}
 \put(3.7,2.7){\framebox(0.6,0.6){${\cal E}_3$}}
  \put(4.3,3){\vector(1,0){0.6}}
  \put(5,3){\text{$\rho^{'}$}}
  \put(6,3){\text{$\Rightarrow$}}

\put(7,2.9){\text{$\rho$}} \put(7.5,2.9){\line(1,0){0.8}}
  \put(6.8,2.2){\text{$|0\rangle_{e_1}$}}
 \put(7.5,2.3){\line(1,0){0.8}}
 \put(8.3,2.2){\framebox(0.8,0.8){$U_1$}}
\put(9.1,2.9){\line(1,0){0.6}} \put(9.1,2.3){\line(1,0){0.6}}

\put(9.7,2.8){\framebox(0.8,0.8){$U_2$}}
\put(9.1,3.5){\line(1,0){0.6}}
  \put(8.4,3.4){\text{$|0\rangle_{e_2}$}}

  \put(10.5,3.5){\line(1,0){0.8}}
   \put(10.5,2.9){\line(1,0){0.8}}

   \put(11.3,2.2){\framebox(0.8,0.8){$U_3$}}
    \put(12.1,2.3){\line(1,0){0.6}}
    \put(12.1,2.9){\line(1,0){0.6}}
    \put(10.7,2.3){\line(1,0){0.6}}
\put(10.1,2.2){\text{$|0\rangle_{e_3}$}}
\put(12.9,2.9){\text{$\rho^{'}$}}

\end{picture}\vskip -15mm
\caption{The left denotes the state evolution of  MO-1gQFA ${\cal
M}$, and the right denotes the resulted machine ${\cal M}^{'}$ that
is to simulate ${\cal M}$. As shown, to simulate the quantum
operation ${\cal E}_i$ in ${\cal M}$, an ancillary quantum system
$E_i$ should be added in ${\cal M}^{'}$ to perform the unitary
operation $U_i$.  This leads to the size of ${\cal M}^{'}$'s
quantum part depending on the length of the input (i.e, the total
running time of quantum operations), which implies that ${\cal
M}^{'}$ is no longer a QFA. }\label{Fig.1}
\end{figure}
We note that, as proved by Aharonov {\it et al} \cite{AKN98},
quantum circuits with mixed states are equivalent to those with
pure states \cite{AKN98}. However, such an equivalence
relationship no longer holds for the restricted model---quantum
finite automata, as we have shown that one-way QFA with mixed
states are more powerful than those with pure states. In fact, the
equivalence between quantum circuits with mixed states and those
with pure states is simply a corollary of such a
 fact that every trace-preserving quantum operation
${\cal E}$ acting on ${\cal H}$ can be simulated by a unitary
transformation $U$ acting on a larger space ${\cal H}\otimes
E$ in such a way ${\cal E}(\rho)={\text
Tr}_E(U\rho\otimes|0^E\rangle\langle0^E|U^{\dagger})$
\cite{AKN98}. Unluckily, such a simulating process is not suitable
for QFA. If we apply this simulating process to mixed-state QFA,
for example MO-1gQFA  ${\cal M}$, and denote the resulted machine by
${\cal M}^{'}$, then as  described in Fig. \ref{Fig.1}, at each
running of a quantum operation  in ${\cal M}$, a new ancillary
quantum system $E$ should be added in ${\cal M}^{'}$. At the same
time,  we know that the total running time of quantum operations
in QFA equals to the length of the input (note that in a quantum
circuit which consists of a finite number of quantum gates and
some input ports,  the total running time of quantum gates has no
dependence on the input). Thus, the resulted machine ${\cal
M}^{'}$ has a quantum part whose size varies with the length of
the input, which is clearly no longer a QFA, since it does not
conform to the definition of QFA given in \cite{MC00,KW97} or any
other we have seen. Likely, quantum finite automata, as a
theoretical model for quantum computers with finite memory, should
have a finite quantum part whose size does not  depend on the
length of the input.

\begin{figure}[htbp]%
 \setlength{\unitlength}{1.3mm}
  \begin{picture}(55,50)
 \put(30,0){\small{MO-1QFA}}
 \put(35,3){\vector(0,1){8}}
 \put(37,3){\vector(1,1){12}}

  \put(32,12){\small{LQFA}}
   \put(36,14){\vector(1,1){12}}
    \put(35,14){\vector(0,1){18}}
     \put(39,13){\vector(2,1){6}}

  \put(45,16){\small{MM-1QFA}}
  \put(49,18){\vector(0,1){8}}

  \put(46,27){\small{GQFA}}
  \put(45,28){\vector(-2,1){8}}

  \put(49,29){\vector(0,1){15}}
  \put(36,36){\vector(3,2){12}}
\put(36,36){\vector(-3,-2){1}}

 \put(30,33){\small{MO-1gQFA}}
 \put(31,32){\vector(-3,-2){12}}
  \put(31,32){\vector(3,2){1}}
 \put(44,45){\small{MM-1gQFA}}

 \put(14,22){\small{ancilla QFA}}
 \put(12,18){\small{Ciamarra QFA}}
  \end{picture}
  \caption{A diagram illustrating known inclusions among the languages recognized with  bounded error by most of the current known 1QFA. Directional lines indicate containments going from the tail to the head; for example, the languages recognized by MO-1QFA are contained in those recognized by  MM-1QFA. Bidirectional lines between two ones mean they contain each other, i.e., they are equivalent; for example, MO-1gQFA and MM-1gQFA recognize the same class of languages. }  \label{Fig.2}
 \end{figure}
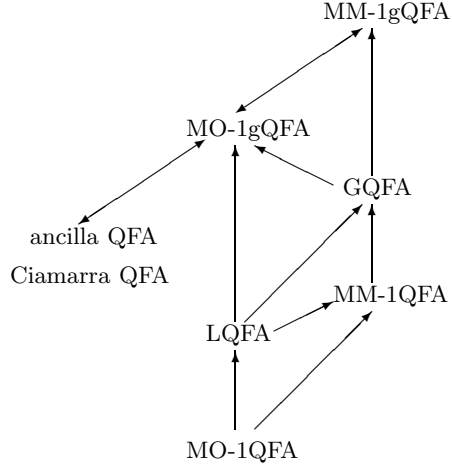

Finally, we present Fig. \ref{Fig.2} to depict the inclusion relations among the languages recognized  by most of the current known 1QFA. Here we use the abbreviations of QFA to denote the classes of languages recognized by them; for example,  ``MM-1QFA'' denotes the class of languages recognized by MM-1QFA with bounder error.  Most of the inclusion relations  depicted in Fig. \ref{Fig.2} are proper inclusions, expect for the following two points: (i) it is still not not known whether GQFA can recognize any language not recognized by MM-1QFA; (ii) MM-1gQFA, MO-1gQFA, ancilla QFA and Ciamarra QFA recognize the same class of languages (i.e., regular languages) as shown in this paper.

\section{ Further discussion }
In this paper, we have addressed the equivalence problem of MO-1gQFA and MM-1gQFA, and  obtained the same equivalence criterion ($(n_1+n_2)^2$, see Theorems \ref{Eqv-MO} and \ref{Eqv-MM}) for both of them. Recently, we noticed that  Ref. \cite{YC10} implied a different method to the equivalence problem of MO-1gQFA and MM-1gQFA, by which the equivalence criterion can be improved to $n^2_1+n_1^2-1$. In fact, this is not an essential improvement. However,  we would like to mention the different method  here, since by comparing the two methods we may have a deeper understanding on QFA.

 From Lemma 1 in \cite{YC10}, we know that  an MO-1gQFA with an $n$-dimensional Hilbert space can be transformed to an equivalent  $n^2$-state {\it Bilinear machine} (BLM) \cite{LQ08a}. In this transformation,  the mapping {\it vec}
 plays a key role. $vec$ is defined as $vec(A)((i-1)n+j)=A(i,j)$ that maps an $n\times n$ matrix $A$ to an $n^2$-dimensional vector. In other words, $vec$ can be defined as $vec(|i\rangle\langle j|)=|i\rangle|j\rangle$.
  An $n$-state BLM  has a form similar to that of probabilistic automata  as shown in Eq.\eqref{PA}: ${\cal A}=(\pi,\Sigma,
\{A(\sigma):\sigma\in\Sigma\},\eta)$, but for BLM, there is no more restriction than that $\pi$ is an $n$-dimensional row  vector, $A(\sigma)$ is an $n\times n$ matrix, $\eta$ is an $n$-dimensional column vector, and all of them have entries in the set of complex numbers. In Ref. \cite{LQ08b}, it was shown that two BLMs with $n_1$ and $n_2$ states, respectively, are equivalent if and only if they are $(n_1+n_2-1)$-equivalent. Therefore, combining the above results, we can obtain the equivalence criterion $n^2_1+n_1^2-1$ for MO-1gQFA.

  Similarly, we can also address the equivalence problem of MM-1gQFA. First, we have proved that an MM-1gQFA can be transformed to an equivalent MO-LM with the same Hilbert space (see Definition \ref{Df-LM} and Lemma \ref{Simulate}). Second, using the mapping $vec$ we can also transform an MO-ML with an $n$-dimensional Hilbert space  to an equivalent  $n^2$-state BLM as did in \cite{YC10}, if only we note that  the  linear super-operator constructed in this paper also has an operator-sum representation as in Eq. \eqref{OP}.

  By the way, in \cite{YC10} it was proved that MO-1gQFA recognize  only stochastic  languages with cut-point.  Based on the results in this paper, we can also prove that MM-1gQFA recognize  only stochastic  languages with cut-point. Thus, MO-1gQFA and MM-1gQFA have the same language recognition power as probabilistic automata  in the sense of both  bounded error and unbounded error.

\end{document}